\documentclass[a4paper,11pt]{article}
\pdfoutput=1 

\usepackage{jcappub} 

\usepackage[english]{babel}
\pdfoutput=1

\usepackage[utf8]{inputenc}
\usepackage[left=2.5cm, right=2.5cm]{geometry}
\usepackage{amsmath,amssymb,mathtools,tabu}
\usepackage{graphicx}
\usepackage{slashed}    
\usepackage{xfrac}                      
\usepackage{url}
\usepackage{color}
\usepackage{placeins}
\usepackage[dvipsnames]{xcolor}
\usepackage{hyperref}
\usepackage{hhline, array,multirow}

\usepackage{float}
\usepackage{caption}
\usepackage{subcaption}
\newcolumntype{P}[1]{>{\centering\arraybackslash}p{#1}}
\newcolumntype{M}[1]{>{\centering\arraybackslash}m{#1}}

\allowdisplaybreaks
\newcommand{\mycite}[1]{\,\cite{#1}}

\newcommand{\lp}[1]{\left(#1\right)}
\newcommand{\mev}{\,\mathrm{MeV}}
\newcommand{\mevsq}{\,\mathrm{MeV}^{-2}}

\newcommand{\ev}{\,\mathrm{eV}}
\newcommand{\invmpc}{\,\mathrm{Mpc^{-1}}}
\newcommand{\hu}{\,\mathrm{km\,s^{-1}Mpc^{-1}}}

\newcommand{\omb}{\Omega_{\rm b}h^2}
\newcommand{\omc}{\Omega_{\rm c}h^2}
\newcommand{\neff}{N_{\rm eff}}
\newcommand{\dNeff}{\Delta N_{\rm eff}}
\newcommand{\geff}{G_{\rm eff}}
\newcommand{\geffa}{G_{\rm eff}^{(1)}}
\newcommand{\geffb}{G_{\rm eff}^{(2)}}
\newcommand{\geffc}{G_{\rm eff}^{(3)}}
\newcommand{\geffi}{G_{\rm eff}^{(i)}}
\newcommand{\lgeff}{{\rm log}_{10}(G_{\rm eff}/{\rm MeV^{-2}})}
\newcommand{\dnu}{\delta_\nu}
\newcommand{\tnu}{\theta_\nu}
\newcommand{\snu}{\sigma_\nu}
\newcommand{\ddnu}{\dot{\delta}_\nu}
\newcommand{\dtnu}{\dot{\theta}_\nu}

\newcommand{\mD}{\mathcal{D}}
\newcommand{\lcdm}{$\Lambda$CDM}
\newcommand{\sinu}{SINU}
\newcommand{\three}{$\mathbf {3c+0f}$}
\newcommand{\two}{$\mathbf {2c+1f}$}
\newcommand{\one}{$\mathbf {1c+2f}$}

\newcommand{\achil}{$\tilde{\chi}_{\ell}^2$}
\newcommand{\achi}{$\tilde{\chi}^2$}

\newcommand{\cb}[1]{\textcolor{black}{#1}}

\abstract{Flavor-universal neutrino self-interaction has been shown to ease the tension between the values of the Hubble constant measured from early and late Universe data. We introduce a self-interaction structure that is flavor-specific in the three active neutrino framework. This is motivated by the stringent constraints on new secret interactions among electron and muon neutrinos from several laboratory experiments. Our study indicates the presence of a strongly interaction mode which implies a late-decoupling of the neutrinos just prior to matter radiation equality. Using the degeneracy of the coupling strength with other cosmological parameters, we explain the origin of this new mode as a result of better fit to certain features in the CMB data.  We find that if only one or two of the three active neutrino flavors are interacting, then the statistical significance of the strongly-interacting neutrino mode increases substantially relative to the flavor-universal scenario. However, the central value of the coupling strength for this interaction mode does not change by any appreciable amount in the flavor-specific cases. We also briefly analyze a scenario with more than three neutrino species of which only one is self-interacting.	In none of the cases, we find a large enough Hubble constant that could resolve the so-called Hubble tension.
	}

\subheader{\hfill SLAC-PUB-17547}

\begin{document}
	\title{\huge{Flavor-specific Interaction Favors Strong Neutrino Self-coupling in the Early Universe}}
	\author[a]{Anirban Das}
	\author[b,c]{and Subhajit Ghosh}
	
	\affiliation[a]{SLAC National Accelerator Laboratory, 2575 Sand Hill Road, Menlo Park, CA 94025, USA.}
	\affiliation[b]{Department of Physics, University of Notre Dame, South Bend, IN 46556, USA.}
	\affiliation[c]{Tata Institute of Fundamental Research, Homi Bhabha Road, Mumbai, 400005, India.}

	\emailAdd{anirband@slac.stanford.edu}
	\emailAdd{sghosh5@nd.edu}
	
	\date{\today}
	\maketitle

	\section{Introduction}
	Neutrino remains to be the most elusive particle in the Standard Model (SM) even after sixty-four years of its discovery. \cb{Even though in the SM, neutrinos are predicted to be massless, the observed phenomena of neutrino oscillation predicts at least one state of the neutrinos to have a mass $ \gtrsim 0.03 \ev$~\cite{nufit,deSalas:2020pgw}. Neutrino oscillation requires  mixing between different flavors, which implies that their mass matrix is non-diagonal in the flavor basis.} Several neutrino oscillation experiments have now measured the neutrino mixing angles to few percent accuracy level. \cb{Terrestrial beta decay experiments probing the absolute masses of neutrinos, however, is yet to reach the sensitivity of that of the oscillation experiments~\cite{Agostini:2019hzm,Adams:2019jhp,KamLAND-Zen:2016pfg}.} 
	In future, beta decay experiment KATRIN is expected to reach a sensitivity of $0.2 \ev$ for electron neutrino mass~\cite{Aker:2019uuj}.
   
	\cb{Cosmology, on the other hand, can probe certain properties of neutrino remarkably better than any terrestrial experiment.} After the advent of state-of-the-art observations of the cosmic microwave background (CMB) by the Planck collaboration, the sum of neutrino masses is now constrained to be below $0.12\,\mathrm{eV}$ \cb{through its dependence on the matter power spectrum\,\cite{Aghanim:2018eyx}. CMB and Big Bang Nucleosynthesis (BBN) measurements of the effective number degrees of freedom in neutrinos in the vanilla \lcdm{} cosmology agrees with the theoretical prediction of the SM. Interestingly, the constraints of neutrino properties from the cosmological data are also very sensitive to the underlying cosmological model. In other words, precision cosmological measurements are sensitive probes for beyond Standard Model (BSM) interactions of neutrinos.} New physics such as a secret interaction among the neutrinos is difficult to probe in terrestrial oscillation experiments because of their very feeble interaction strength.  
	\cb{Whereas, the dense environment in the early Universe can amplify the effects of any secret interaction in the neutrino sector due to large number density.}

	Several BSM
	interaction scenarios have been invoked in the neutrino sector to explain several experimental measurements and observations, e.g., anomalous neutrino signal in short-baseline experiments\mycite{Aguilar:2001ty,Aguilar-Arevalo:2018gpe}, discrepancy between the CMB and local measurement of the Hubble parameter\mycite{Aghanim:2018eyx,2018ApJ...855..136R} etc. One of the proposals to address these observations/tensions is self-interacting neutrino (\sinu)\mycite{Hannestad:2013ana, Dasgupta:2013zpn,Archidiacono:2014nda,Chu:2015ipa,Cyr-Racine:2013jua,Lancaster:2017ksf,Oldengott:2017fhy,Ghosh:2019tab}. The deviation of the neutrino sector from its vanilla free-streaming nature have been shown to affect various cosmological observables. In the early universe, the dense neutrino environment gives us an opportunity to test these \sinu{} models\mycite{Duan:2005cp,Duan:2006an,Duan:2006jv}. Several studies have been done on the effects of self-interaction among neutrinos on CMB and large scale structure\,\cite{Trotta:2004ty,Hannestad:2004qu,Bell:2005dr,Melchiorri:2006xs,Friedland:2007vv,DeBernardis:2008ys,Gerbino:2013ova,Archidiacono:2013fha,Melchiorri:2014yoa}.  Most of these studies are based on a parametrization of the self-interacting neutrino sector using $c_\mathrm{eff}$ and $c_\mathrm{vis}$. However, with the availability of more precise experimental data, it is now important to model the \sinu{} using a realistic particle physics-based approach. Previous studies have also shown that the neutrino flux from a supernova will be sensitive to new interactions among the neutrinos\mycite{Blennow:2008er,Das:2017iuj}.
	\cb{However, the strongest constraints on the parameter space of the \sinu{} models come from $K$-meson decay, double beta decay, invisible width of the $Z$-boson, and $\tau$-decay experiments\mycite{BILENKY1993287,Blinov:2019gcj,Brdar:2020nbj,Lyu:2020lps}.} Therefore, modelling a BSM \sinu{} scenario that can address the above anomalies, while respecting the phenomenological constraints, is a challenging task~\cite{He:2020zns,Berbig:2020wve}. 
	
	\begin{figure}[t]
		\begin{center}
			\includegraphics[width=0.5\linewidth]{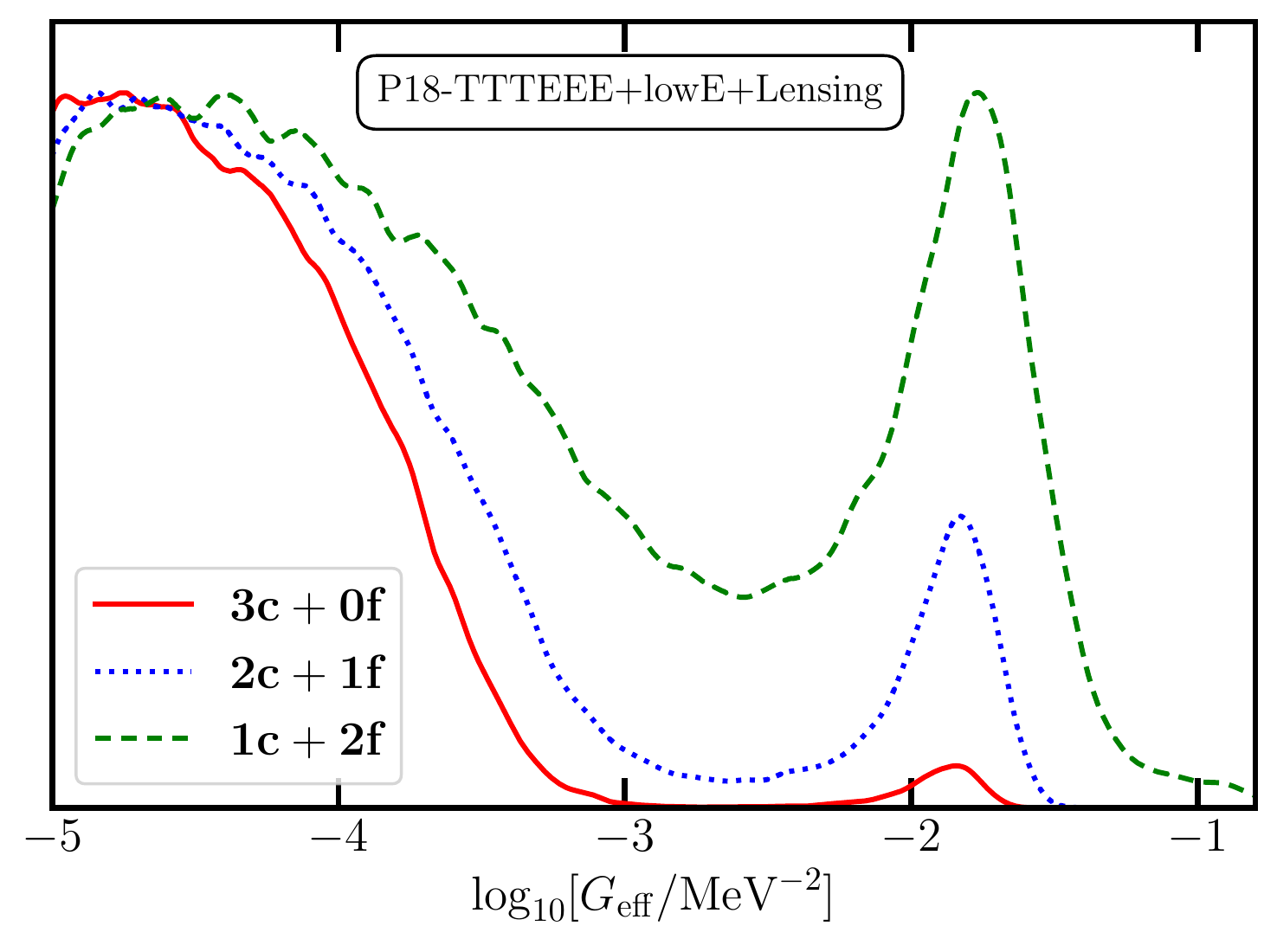}
			\caption{The 1D posteriors for $\lgeff$ in the three scenarios- \three{}, \two{}, and \one{} with three, two, and one self-interacting neutrinos, respectively. The two-mode feature is present in all cases. The relative significance of the larger $\geff$ mode increases with less number of interacting neutrinos, hence, is largest in the \one{} scenario. 
				 See section\,\ref{sec:result} for more details.}
			\label{fig:1d_posterior}
		\end{center}
	\end{figure}

	As alluded earlier, self-interacting neutrinos have been suggested to ease a \cb{crisis of modern cosmology} - the so-called Hubble tension. Since the early days of the Planck experiment, a \mbox{discrepancy} in the value of the Hubble parameter from CMB observations and local (low redshift) measurements have become a topic of active research\mycite{Ade:2013zuv}. This discrepancy has been exacerbated with the latest data from the Planck experiment and the local $H_0$-measurement using cepheids and supernovae\mycite{Aghanim:2018eyx,Verde:2019ivm,Riess:2019cxk,2018ApJ...855..136R}. The Planck-measured value of the Hubble constant using the temperature, polarization and lensing CMB data is $H_0=67.36\pm 0.54\hu$, whereas the SH0ES collaboration found $H_0=74.03\pm 1.42\hu$. There is now a $4.4\sigma$ tension between the two values of $H_0$.
	However, an independent local measurement of $H_0$ was done in Ref.\cite{freedman2019} which found a value $H_0=69.8\pm 1.9\hu$. Apart from unknown experimental systematics, this discrepancy could point towards a new physics signature beyond the \lcdm{} paradigm. 
	Several such scenarios have been envisaged to address this tension \mycite{DiValentino:2016hlg,Qing-Guo:2016ykt,Kumar:2016zpg,Renk:2017rzu,DiValentino:2017zyq,DiValentino:2017rcr,DiValentino:2017oaw,Bolejko:2017fos,Khosravi:2017hfi,DEramo:2018vss,Dutta:2018vmq,Banihashemi:2018has,belgacem2018,Pandey:2019plg,Agrawal:2019lmo,Agrawal:2019dlm, DiValentino:2019exe,Desmond:2019ygn, Pan:2019cot,Vattis:2019efj,Poulin:2018cxd,lin2019,lin2019b,Cyr-Racine:2013jua,Lancaster:2017ksf,Kreisch:2019yzn,Karwal:2016vyq,Ghosh:2019tab}. Among them, Ref.\mycite{Cyr-Racine:2013jua,Lancaster:2017ksf} introduced new \emph{secret} interaction between Standard Model active neutrinos mediated by a new heavy scalar $\phi$.  A flavor-universal four-Fermi interaction with a coupling strength $\geff$ was assumed,
	\begin{equation}\label{eq:geff1}
	\mathcal{L} \supset \geff\bar{\nu}\nu\bar{\nu}\nu, \qquad \geff \equiv \frac{g_\phi^2}{M_\phi^2}\,,
	\end{equation}
	where $g_\phi$ is the coupling between $\nu$ and the mediator $\phi$, and $M_\phi$ is the mass of $\phi$. A Bayesian analysis of this model with the CMB data prefers two distinct regions of $\geff$ values: strongly interacting (SI) mode with a large value $\lgeff \simeq -1.711^{+0.099}_{-0.11}$ ($68\%$ confidence limit), and the moderately interacting (MI) mode with an upper bound $\lgeff < -3.57$ at $95\%$ confidence level using the Planck 2015 and BAO data. Note that the SI mode coupling strength is about $10^{9}$ times stronger than the weak interaction Fermi constant $G_F = 1.17\times 10^{-11}\mevsq$. 

	\cb{However, \sinu{} scenario meets strong constraints from the laboratory experiments \mycite{Blinov:2019gcj,Lyu:2020lps}. Ref.\mycite{Blinov:2019gcj} showed that a simple model of flavor-universal \sinu{} scenario is ruled out by laboratory constraints from meson decay, $ \tau $ decay, and double beta decay. Whereas, Ref.\mycite{Lyu:2020lps} demonstrated that Standard Model Effective Field Theory (SMEFT) operators, giving rise to flavor universal \sinu{} scenario, are also highly constrained by meson decays, $ Z $ decays and electroweak precision measurements etc. These studies also derive flavor-dependent \sinu{} constraints where only one of the neutrinos is interacting. The constraints in those cases are very stringent for $ \nu_e $, whereas for $ \nu_\mu $ and $ \nu_\tau $ the constrains are comparatively weaker.}
	
	\cb{In this paper, we study flavor-specific neutrino self-interaction scenario using latest cosmological data.
	We perform a detailed Bayesian analysis of \sinu{} with flavor-specific interactions for the first time using the latest 2018 likelihood from the Planck collaboration. As such, we allow the coupling strengths for different neutrino flavors to be different from each other. Our goal is to complement the flavor-specific \sinu{} studies from laboratory experiments using the cosmological data. However, there is a notable distinction between the framework of \sinu{} studies in cosmology vs laboratory searches. In laboratory experiments, the flavor of neutrino can be identified via electroweak interaction. 
	Whereas, cosmological observables such as CMB are only sensitive to the temperature, free-streaming properties and the total mass of the neutrinos. Therefore, in cosmology, we cannot distinguish any particular flavor of neutrino and can only study their effects collectively. 
	All three generation of neutrinos $ \nu_e, \nu_\mu $ and $ \nu_\tau $ are on equal footing for our analysis. }
	
	\cb{
	In this work, we considered only three massless SM neutrinos, and therefore, fixed $ N_{\rm eff}  = 3.046 $ ($ N_{\rm eff} \approx 1.015 $ for each flavor). Massless neutrinos are a fairly good approximation as the bound on the total mass of the neutrinos are an order of smaller compared to the temperature of the plasma at the last scattering surface. The effect of massive neutrinos is reasonably well understood and can be speculated from our results. We consider
	three possible scenarios depending on the interactions of the three neutrino species -- 3-coupled (\three), 2-coupled + 1-free-streaming (\two), and 1-coupled + 2-free-streaming (\one) respectively. We assumed same coupling strengths for the coupled species for the first two cases. This is justified because the coupled neutrino species are completely equivalent due to the massless approximation. 
	Setting neutrino mass to zero implies that the neutrinos remain in the same flavor eigenstate while propagating and, hence, there is no mixing between different flavor eigenstates. Note that, the phenomenological analysis performed here is also applicable to the scenarios where neutrinos have tiny masses, hence non-zero mixing, but the interaction is diagonal in the mass basis. However, in this paper, we treat neutrinos to be \emph{exactly} massless and thus do not consider mixing.
	As a result, the flavor structure of the interaction remains the same during propagation. Also note that, the \three{} case is identical to the universal flavor coupling scenario which has been studied previously. Additionally, we also analyze the scenario \one$+\dNeff$, where the total $\neff$ is allowed to vary, in appendix\,\ref{sec:app3}}.

Our main result is shown in figure\,\ref{fig:1d_posterior}. We find that the flavor-universal case yields a low significance for the SI mode with the Planck 2018 data. However, the SI mode significance is drastically increased once the flavor-universality of the coupling strength is relaxed, becoming maximum when only one neutrino state is self-interacting. We also find that the SI mode significance even surpasses the MI mode in certain cases. 
We explain the origin of the SI mode as a results of a better fit of certain features of the CMB data, compared to \lcdm{}, using degeneracy of $\geff$ with other cosmological parameters.
When only one or two neutrino states are self-interacting, the resulting changes are smaller compared to the scenario when all three neutrinos are interacting and hence, can be compensated by the other correlated parameters relatively easily. This results in substantial enhancement of the significance of SI mode in \two{} and \one{} cases. We also find that the SI mode best-fit value of $\geff$ has a mild dependence on the number of interacting species.

	The outline of the paper is as follows. In section \ref{sec:setup}, we describe our cosmology model and explain the rationale behind it. In section \ref{sec:method} we detail our methodology and the experimental likelihoods used in this work. We show and interpret our results in section \ref{sec:result}, and conclude with a discussion in section \ref{sec:discussion}.
	
	\section{Setup}\label{sec:setup}
	\subsection{The model}\label{sec:model}
	We consider scalar interactions between massless $\nu$ and $\phi$ \cb{below the electroweak scale }as
	\begin{equation}
	\mathcal{L} \supset g_{ij}\phi\bar{\nu}_i\nu_j\,,
	\end{equation}
	where $g_{ij}$ is the coupling between $\phi$ and the neutrino flavors $i$ and $j$\footnote{Note that the mediator could also be a vector particle which would change the details of the interaction, but the phenomenological aspect of the model remains the same.}. When the temperature of the neutrino bath $(T_\nu)$ cools down below the mass of $\phi$, i.e., $T_\nu \ll M_\phi$, a four-Fermi interactions among the neutrinos, similar to Eq.(\ref{eq:geff1}), is generated as follows
	\begin{equation}\label{eq:geff2}
	\mathcal{L} \supset \geff^{(ijkl)}\bar{\nu}_i\nu_j\bar{\nu}_k\nu_l, \qquad \geff^{(ijkl)} \equiv \frac{g_{ij}g_{kl}}{M_\phi^2}\,.
	\end{equation}
	Here we note that for a general interaction of $\phi$ with any two flavors of neutrino, the four-Fermi interaction strength $\geff^{(ijkl)}$ in Eq.(\ref{eq:geff2})  has four indices for a process like $\nu_i + \nu_j \to \nu_k + \nu_l$. 
	Therefore, the most general scenario would involve many different couplings for different flavor combinations.
	 In such a case, the energy and momentum of individual neutrino species will not be conserved and one will need to incorporate energy and momentum transfer between different species in the perturbation equations\,\cite{Oldengott:2017fhy}. However, in this work we limit ourselves to a simpler scenario as described below. We consider only \emph{diagonal} interactions 
	  in the flavor space with different coupling strengths:
	  \begin{equation}\label{eq:eff2diag}
	  \mathcal{L} \supset \geffi \bar{\nu}_i\nu_i\bar{\nu}_i\nu_i\;.
	  \end{equation}
	   This assumption introduces only three new \sinu{} parameters: $\geffa,\ \geffb$ and $\geffc$. 
	   As mentioned earlier, in this work we fix the number of neutrino flavors to three with all of them having a same temperature $T_\nu$. However, because of the complete equivalence of the interacting states in the context of CMB, we need to consider only one common coupling parameter $\geff$ for all the interacting states for a given scenario.

	We also assume $M_\phi > 1\mev$ to avoid BBN constraints on extra relativistic species around \cb{plasma temperature} $T\sim 1\mev$\mycite{2011PhLB..701..296M,Cooke:2013cba}. The four-Fermi interaction picture is valid only below $T_\nu=M_\phi$ when $\phi$ cannot be produced from scattering of the neutrinos. 
	However for $M_\phi>1\mev$, 
	its population is Boltzmann suppressed
	and hence does not affect the evolution of the density fluctuations during the period $ (T < 100 \ev)$ relevant to the present analysis. The annihilation and decay of $\phi$ into neutrino increases the temperature of the neutrino bath, but this is model-dependent and we do not take this extra heating into account in this work.\footnote{Although, we note that in a concrete model, the heating due to $\phi$ leads to a larger $\neff$ which could help increase the Hubble parameter to some extent.}
	\begin{figure}[t]
		\begin{center}
			\includegraphics[width=0.7\linewidth]{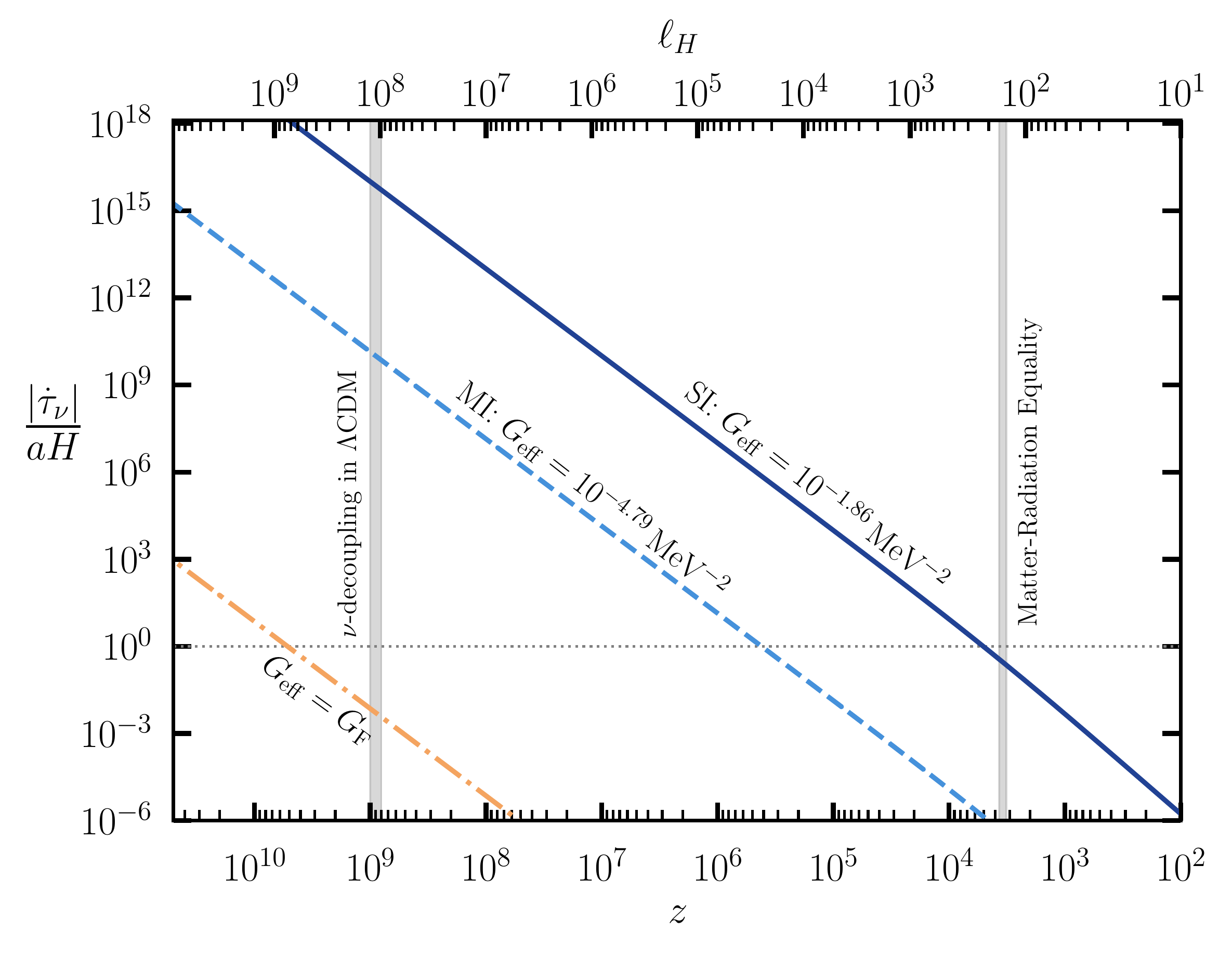}
			\caption{Variation of the neutrino self-interaction opacity $\dot{\tau}_\nu$ relative to the Hubble expansion rate $H$ with redshift for $\geff= 10^{-1.86}\mev^{-2}$ (dark blue, solid), $10^{-4.79}\mev^{-2}$ (light blue, dashed), and $G_\mathrm{F}$ (light brown, dot-dashed). The chosen values of $\geff$ are the best-fit values for SI and MI mode in \three{} scenario with Planck temperature and polarization data. The top axis shows the angular multipole $\ell_H$ corresponding to the modes that enter horizon at redshift $ z $.
				The epochs of neutrino decoupling in \lcdm{} and the matter-radiation equality are shown as gray-shaded region.}\label{fig:opacity}
		\end{center}
	\end{figure}
	
	Using dimensional analysis, the thermally averaged scattering cross section between the $i$-th neutrino state goes as $\langle\sigma v\rangle \sim (\geff)^2T_\nu^2$. Therefore the interaction rate scales as \mbox{$\Gamma_\nu \equiv n_\nu\langle\sigma v\rangle$} $\sim (\geff)^2T_\nu^5$ because of the $T_\nu^3$-scaling of the neutrino number density. We absorb all other model-dependent prefactors of the interaction rate into $\geff$. The \cb{comoving} neutrino self-interaction opacity $\dot{\tau}_\nu$ is defined as
	\begin{equation}\label{eq:opacity}
	\dot{\tau}_\nu = -a(\geff)^2T_\nu^5\,,
	\end{equation}
	where $a$ is the scale factor of the Universe. Note that the $T_\nu^5$-scaling of the opacity is a characteristic of the four-Fermi interaction. The neutrinos are self-interacting when the interaction rate $\dot{\tau}_\nu$ is greater than the comoving Hubble expansion rate $aH$, i.e., $\dot{\tau}_\nu > aH$. The interaction freezes-out when $\dot{\tau}_\nu$ drops below the Hubble expansion rate at a redshift $z_\mathrm{dec}$ which is given by 
	\begin{equation}
	1+z_\mathrm{dec} \simeq 1.8\times 10^4 \left(\frac{\geff}{10^{-2}\mev^{-2}}\right)^{-\frac{2}{3}}.
	\end{equation}
	In the above equation, we have set other background cosmological parameters to their \lcdm{} best-fit values and assumed the decoupling to take place in radiation domination era.
	We show the relative strength of the neutrino self-interaction to the Hubble expansion rate in figure \ref{fig:opacity}. From figure \ref{fig:opacity}, we see that the neutrinos are interacting with each other until 
	$z \simeq 4000  $ for $ \geff = 10^{-1.86} \mev^{-2}$ which is much later than SM neutrino decoupling $ (z \simeq 10^9) $. For the MI modes the decoupling happens at $ z > {\rm a ~few } \times 10^5$.
	\cb{The neutrino decoupling for the SI mode happens very close to the matter-radiation equality which has interesting implications that will be discussed later.}
	
	Note that, because of the massless approximation, the structure of the interaction matrix does not change during propagation. However, this will not be true in case of massive neutrinos because of mixing between different flavor eigenstates. In that case, interaction of a particular flavor will generate interaction among different mass eigenstates depending on the structure of the mixing matrix. However, in our case, there is no mass mixing and the flavor eigenstates are the eigenstates of the Hamiltonian due to massless approximation.
	
   \subsection{Perturbation equations}\label{sec:perturb}
   The Boltzmann hierarchy of the perturbation equations for massless, self-interacting neutrinos in the Newtonian gauge is shown below following Refs.\mycite{Ma:1995ey,Cyr-Racine:2013jua}.
   \begin{equation}\label{eq:boltzmann_hierarchy}
   \begin{array}{l}
   \ddnu + \dfrac{4}{3}\tnu - 4\dot{\phi} = 0\,,\\[2ex]
   \dtnu + \dfrac{1}{2}k^2\lp{ F_{\nu,2} - \dfrac{1}{2}\dnu} - k^2\phi = 0\,,\\[2ex]
   \dot{F}_{\nu,\ell} + \dfrac{k}{2\ell+1} \lp{(\ell+1)F_{\nu,\ell+1} - \ell F_{\nu,\ell-1}} = \alpha_\ell \dot{\tau}_\nu F_{\nu,\ell},\quad \ell \geq 2\,.
   \end{array}
   \end{equation}
   Here $\alpha_\ell$ are $\ell$-dependent $\mathcal{O}(1)$ angular coefficients 
   that depend on the details of the neutrino interaction model, and the anisotropic stress $\snu$ is related to $F_{\nu,2}$ as $\snu = F_{\nu,2} / 2$. 
   Energy and momentum conservation dictates $\alpha_0=\alpha_1=0$. The values of $\alpha_\ell$ for $\ell \ge 2$ is approximately of order unity. In this work, we take $\alpha_\ell=1$ for $\ell \ge 2$ which is a fairly good approximation.  We implement these new perturbation equations for each neutrino species in the public code \texttt{CLASS} and solve them \cb{numerically}\mycite{2011JCAP...07..034B,Archidiacono:2019wdp}.\footnote{The modified \texttt{CLASS} code is available at \href{https://github.com/anirbandas89/CLASS_SInu}{https://github.com/anirbandas89/CLASS\_SInu}.} For very large value of the coupling \cb{$ \geff $} when $|\dot{\tau}_\nu| \gg aH$, this system of equations may become difficult to solve as the equations become stiff, and tight-coupling approximation may be necessary. However, we checked that the default \texttt{ndf15} integrator in \texttt{CLASS} is able to solve the equations for $\lgeff \leq -0.8$ without invoking tight-coupling approximation which turns out to be sufficient for this work.
   
   
   In the presence of self-interaction, neutrinos are strongly coupled at very early time which modifies the initial conditions for the perturbation variables compared to \lcdm. For strongly coupled fluids, the initial anisotropic stress is zero. However, for large scale modes entering the horizon after neutrino decoupling this characterization is not true. For those modes, the initial anisotropic stress is nonzero as neutrinos are free-streaming when the modes enter horizon. To set unique initial conditions for all the modes of interest of CMB, we start evolving each $ k $-mode in CLASS from a very high redshift $ (z\sim 10^7) $. Because at $ z \sim 10^7 $, $ |\dot{\tau}_\nu| \gg (aH) $ for all SINU modes including the MI mode, we can safely set initial anisotropic stress to zero for \emph{all} $ k $ values. We also modify other initial conditions accordingly.
   	However, we found that the resulting SINU spectrum with these modified initial conditions differ very slightly $ (\lesssim 0.1\%) $ compared to SINU model with \lcdm{} initial spectrum where initial anisotropic stress of neutrinos is non-zero. This is because, the anisotropic stress for a $k$ mode, that starts with a non-zero value, vanishes very quickly due to the strong self-interaction.
   	Since the modifications due to the initial conditions are very small compared to the precision of the Planck data and also makes the code comparatively much slower due to the additional integration time, we chose not to incorporate those for the mcmc analysis. 
   
   \begin{figure}[t]
   	\begin{center}
   		\includegraphics[width=\linewidth]{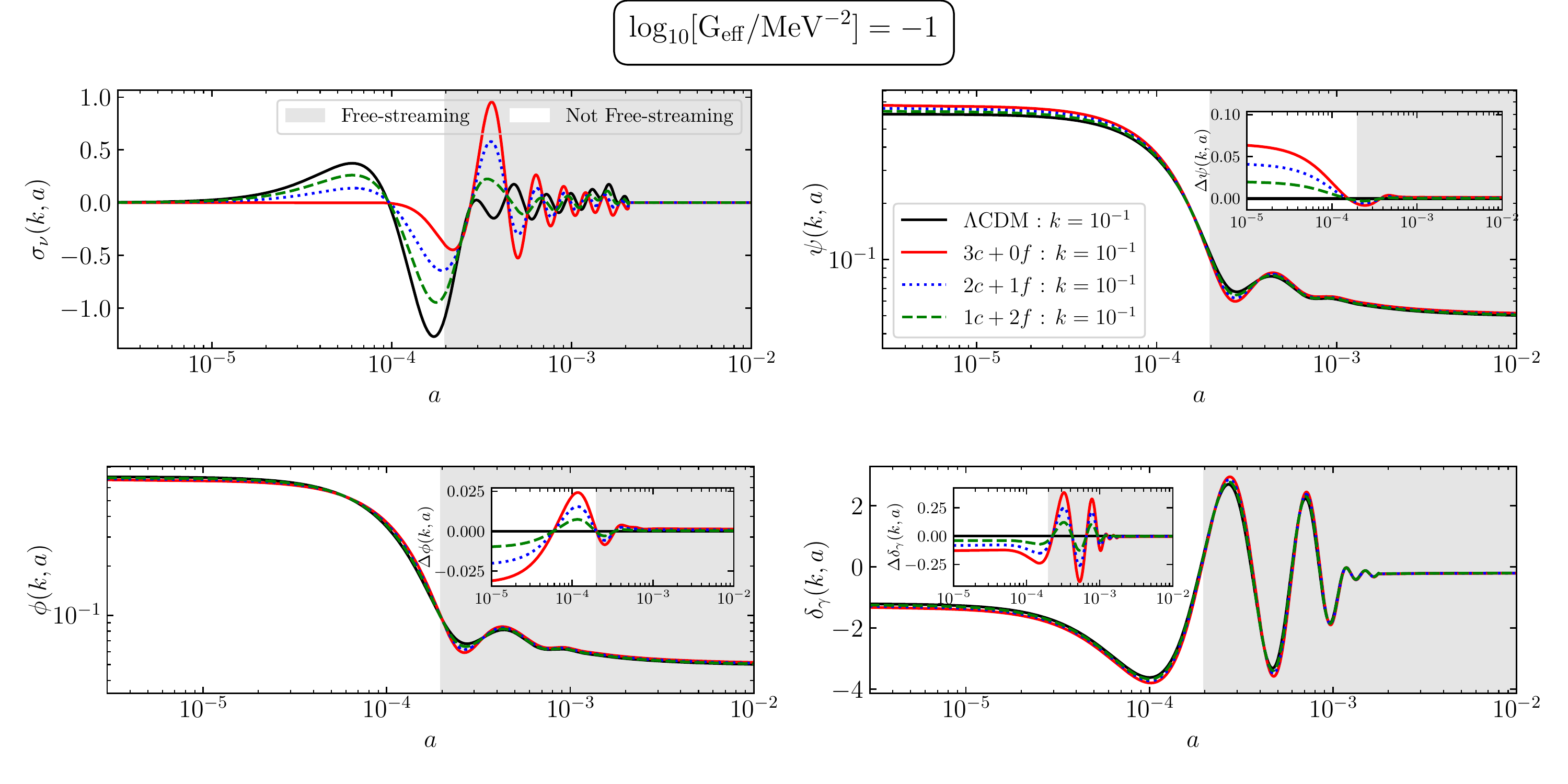}
   		\caption{Evolution of the neutrino anisotropic stress $\sigma_{\nu}$ (top left), the gravitational potentials $ \psi $ (top right), $ \phi $ (bottom left), and the photon over-density $ \delta_\gamma $ (bottom right) for the mode $k=0.1\invmpc$  for \three{} (red, solid), \two{} (blue, dotted), \one{} (green, dashed), and  \lcdm{} (black, solid). The white region in each plot, which is determined by $ |\dot{\tau}_\nu|/(aH) \ge 10 $, is approximately the region in scale factor $ a $ upto which neutrinos are tighly-coupled for $ \geff = 0.1 \mev^{-2} $. From the top left plot we see that the $ \sigma_{\nu} $ is suppressed in the non-free streaming region due to the self-interaction. The suppression is largest for \three{} where all three neutrinos are tightly coupled, and gradually decreases for \two{} and \one{}, respectively. The suppression of $ \sigma_{\nu} $ results in the enhancement of $ \phi $ and $\psi$ which in turn enhances $ \delta_\gamma $. The insets show the absolute changes in the corresponding cases compared to \lcdm{}.}
   		
   		\label{fig:shear_evol}
   	\end{center}
   \end{figure}


   \subsection{Changes in the CMB power spectra}\label{sec:changes_in_CMB}
   In this section, we shall discuss the changes in the CMB angular power spectra due to neutrino self-interaction.
   The self-interaction stops the neutrinos from free-streaming before decoupling. As can be seen from Eq.(\ref{eq:boltzmann_hierarchy}), the new interaction plays the role of \emph{damping} in the perturbations for $\ell \ge 2$. Therefore, it impedes the growth of the anisotropic stress $\sigma$ while the neutrinos are strongly-coupled. We show the evolution of neutrino anisotropic stress $\snu$ in the top left panel of figure\,\ref{fig:shear_evol} in both \sinu{} and \lcdm{} cosmology. The important difference between  them is the initial suppression of $F_{\nu,2}$ in the \sinu{} scenario caused by the new interaction. \cb{The suppression is maximum for \three{} where all three neutrinos are interacting, and gradually decreases for \two{} and \one{} where the number of interacting neutrino flavor is two and one respectively. }The anisotropic stress is related to the gravitational potentials $\phi$ and $\psi$ via the following equation,
   \begin{equation}
   k^2(\phi-\psi) = 12\pi Ga^2\sum_{i=\gamma,\nu}(\rho_i+P_i)\sigma_i\, \simeq 16\pi Ga^2\rho_{\rm tot}R_\nu\sigma_\nu\,,
   \end{equation}
   where $\rho_i, P_i, \sigma_i$ are individual energy density, pressure and anisotropic stress, respectively, for the $i$-th species, and $\rho_{\rm tot}$ is the total energy density. In the last step of the above equation, we have ignored the small anisotropic stress of photon $\sigma_\gamma$ before recombination. Also, $R_\nu$ is the fractional energy density of \emph{free streaming} neutrinos which, in radiation domination, is given by
   \begin{equation}\label{eq:Rnu}
   R_\nu = \dfrac{\rho_\nu}{\rho_\nu + \rho_\gamma}\;.
   \end{equation}
   In the radiation domination era $R_\nu \sim 0.41$ in \lcdm{}. Therefore, the suppression of neutrino anisotropic stress due to \sinu{} during this period plays an important role in enhancing the gravitational potentials $\phi$ and $\psi$, as can be seen from figure~\ref{fig:shear_evol}. The gravitational potentials in turn affect the evolution of the photon perturbations as can be seen in the bottom-right panel of figure\,\ref{fig:shear_evol}.
      \begin{figure}[t]
   	\centering
   	\includegraphics[width=1.0\linewidth]{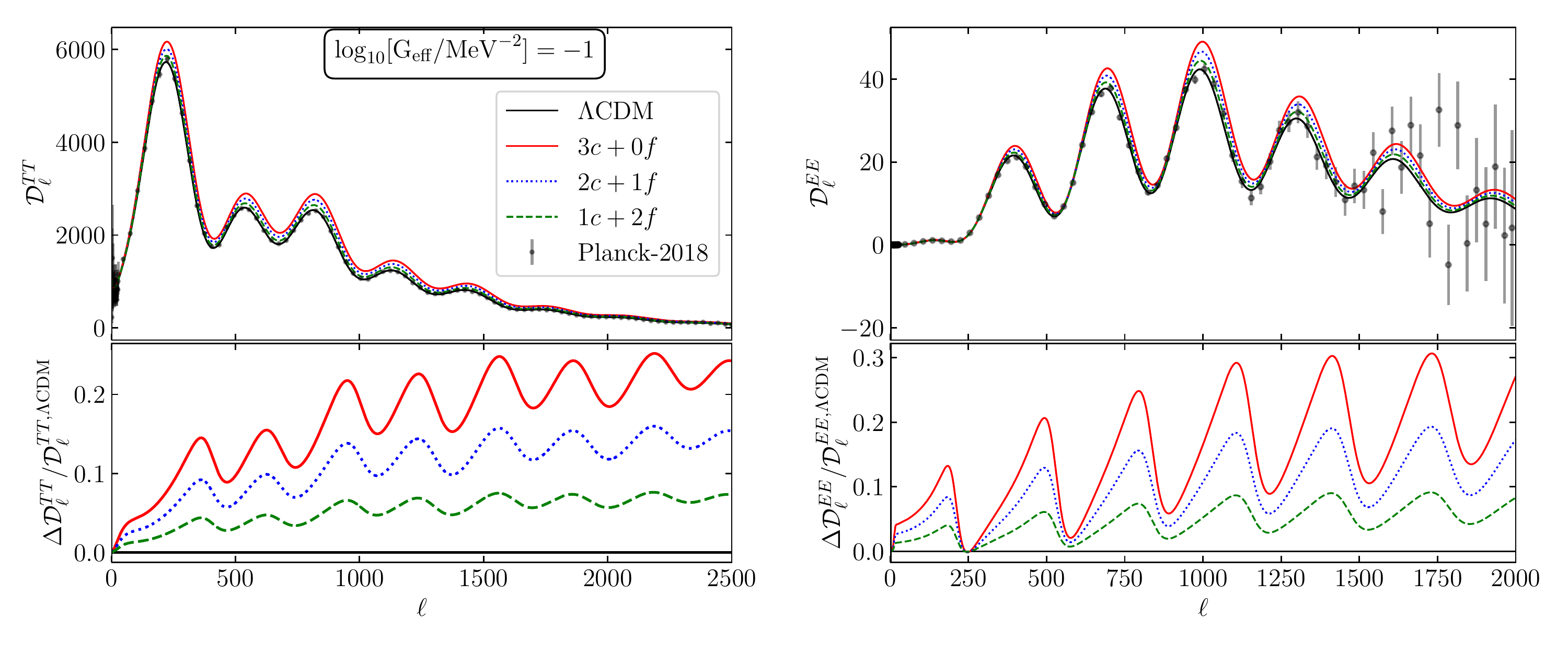}
   	\caption{CMB $ TT $ and $ EE $ angular power spectra for \three{} (red, solid), \two{} (blue, dotted), and \one{} (green, dashed) scenarios  are shown in the top panels. The \lcdm{} power spectra are also shown for comparison in solid black. The parameters for the \lcdm{} spectra correspond to the best-fit points for the TT,TE,EE+lowE dataset. The bottom panels show the relative changes from the \lcdm{} spectra. For \sinu{} plots, we have set $ \lgeff =-1 $ and the rest of the parameters are fixed to their \lcdm{} best-fit values. We also show the binned Planck 2018 data in both the plots as black circles with errorbar. The enhancement and the phase shift of the CMB spectra are evident from the bottom panels. Both the effects increase with the number of self interacting neutrino species.}\label{fig:cmb_spectra}
   \end{figure}

  The changes in the CMB power spectrum in figure.~\ref{fig:cmb_spectra} can be understood
  as a result of change in the propagation speed for perturbation of the neutrinos as explained below following Refs.\,\cite{Bashinsky:2003tk,Baumann:2015rya}. In \lcdm{} cosmology, the  neutrino perturbations travel at the speed of light while the perturbations in the photon-baryon fluid propagates at the speed of sound $c_s \simeq 1/\sqrt{3}$. Therefore, the neutrinos tend to create perturbations ahead of the  sound horizon. This creates a \emph{phase shift} $\phi_\nu$ in the acoustic oscillations of the photon\,\cite{Bashinsky:2003tk}\,, which in the radiation domination, is given by
  \begin{equation}\label{eq:phase_shift}
  \phi_\nu \simeq 0.19\pi R_\nu\,.
  \end{equation}
  The amplitude of the oscillation is also changed by a factor of $(1+\Delta_\nu)$ where
  \begin{equation}\label{eq:amp_supp}
  \Delta_\nu \simeq -0.27 R_\nu\,.
  \end{equation}
   This results in a suppression of the CMB power spectrum in \lcdm{} due to free streaming neutrinos for the modes that enter the horizon before matter-radiation equality. Therefore, these changes in the photon acoustic oscillations affect the CMB angular power spectra for $\ell \gtrsim 200$. Effectively, the phase shift moves the peaks in the TT and EE power spectra towards smaller $\ell$, and suppresses the power spectra~\cite{Bashinsky:2003tk}.
  
  This story is changed in the presence of self-interacting neutrinos. The self-interaction stops the neutrinos from free-streaming and delays the neutrino decoupling from the thermal bath until a later time $(z_{\rm dec})$ depending on the strength of the interaction. As a result, the free-streaming neutrino fraction $R_\nu$ is decreased relative to its \lcdm{} value depending on the number of neutrino species which are coupled at a certain time:
  \begin{equation}\label{eq:rnuval}
  R_\nu = R_\nu^\text{\lcdm{}} \times \begin{cases}
  0,~~\,\qquad {\rm for~}\text{\three}\\[1.5ex] {\Large \sfrac{1}{3}},\qquad {\rm for~}\text{\two} \\[1.5ex] {\Large\sfrac{2}{3}},\qquad {\rm for~}\text{\one}
  \end{cases}
  \end{equation}
  Decreasing $R_\nu$ in Eq.(\ref{eq:phase_shift}) and (\ref{eq:amp_supp}) readily implies a phase shift of the CMB spectrum towards larger $\ell$, and an enhancement of power relative to \lcdm. The resulting changes in the CMB \cb{TT and EE} spectra can be seen in figure\,\ref{fig:cmb_spectra}. In the upper panels, we show $\mD_\ell^{XX}=\ell(\ell+1)C_\ell^{XX}/(2\pi)$ where $XX=TT~\text{and}~EE$ in three \sinu{} scenarios and \lcdm{}. In the lower panels, we show the fractional changes of the spectrum relative to \lcdm{}. We see that there is an enhancement in the \sinu{} spectra, and also a phase shift which shows up as wiggles in the fractional difference plot. In compliance with the explanation above, both of these effects are maximal in \three{} and gradually decreases as the number of self interacting states is decreased. Taking everything into account, we can see that the overall changes in the spectrum are milder when less number of neutrinos are interacting, which allow these changes to be compensated relatively easily by changing other parameters.
  
  A higher Hubble parameter can undo the change in the neutrino induced phase shift $\phi_\nu$ in \sinu{}.
  This can be understood in terms of the photon transfer function $\cos(kr_s^\ast + \phi_\nu)$ which sources the acoustic peaks in the CMB spectrum. The CMB multipole corresponding to a mode $k$ is given by,
  \begin{equation}\label{eq:lpeak}
  \ell \approx k D_A^* = (m\pi - \phi_\nu)\dfrac{D_A^\ast}{r_s^\ast}
  \end{equation}  
  where $D_A^\ast $ is the angular diameter distance, and $r_s^\ast$ is the sound horizon at recombination defined as,
  \begin{equation}\label{eq:rsandda}
  D_A^\ast = \int_{0}^{z^\ast}\dfrac{1}{H(z)} dz , \qquad r_s^\ast = \int_{z^\ast}^\infty\dfrac{c_s(z)}{H(z)} dz
  \end{equation} where $H(z) $ is the Hubble rate, and $c_s(z) \approx 1/\sqrt{3}$ is the sound speed in the photon-baryon fluid.
  Decrease in the phase shift $\phi_\nu$ from its \lcdm{} value moves the spectrum towards a higher $\ell$ value as can be seen from Eq.\eqref{eq:lpeak}, which can be compensated by an increase in $\theta_\ast \equiv  r_s^\ast / D_A^\ast $. This shift in $\theta_\ast$ can be accommodated by changing $D_A^\ast$ via modifying $H_0$ and $\Omega_\Lambda$ which changes the Hubble evolution at late times~\cite{Ghosh:2019tab}. The increase in Hubble constant $H_0$ is only relevant for SI mode where the self interaction strength is significantly strong inducing a large phase shift. In other words, higher values $\geff$ is positively correlated with $H_0$. Note that, the increase of $H_0$ for SI mode is maximum in \three{} case where all three neutrinos interact producing largest phase-shift compare to \lcdm{}.

\begin{table}[t]
	\centering
	\caption{The nested sampling settings used in this work.}\label{tab:ns}
	\begin{tabular}{M{4cm}M{2cm}}
		\hline\hline
		Parameter & Value\\ \hline
		Sampling efficiency & $0.8$ \\
		Evidence tolerance & $ 0.1 $\\ 
		Live points & $2000$\\
		\hhline{--}
	\end{tabular}
\end{table}

   	\begin{table}[b]
  
   		\caption{Prior ranges used for all parameters except $\lgeff$.}\label{tab:prior}
   		\centering
   		\begin{tabular}{M{2.4cm}M{3.5cm}}
   			\hline\hline
   			Parameter & Prior\\[1ex] \hline
   			$\omb$ & $[1.00, 4.00]$\\ [1ex]
   			$\omc$ & $[0.08, 0.16]$\\[1ex]
   			$100\theta_s$ & $[2.00, 4.00]$\\[1ex]
   			$\tau_\mathrm{reio}$ & $[0.004, 0.25]$\\[1ex]
   			$\ln(10^{10}A_s)$ & $[2.00, 4.00]$\\[1ex]
   			$n_s$ & $[0.90, 1.02]$\\[1ex]
   			\hhline{--}
   		\end{tabular}
 \end{table}

	\section{Data \& Methodology}\label{sec:method}
	We performed a Bayesian analysis of this model using the latest version of the Markov Chain Monte Carlo (MCMC) sampler \texttt{MontePython3.3}\mycite{Audren:2012wb,Brinckmann:2018cvx}. To analyze the MCMC chains and plot the parameter posteriors, we use the \texttt{GetDist 1.1.2} software package\,\cite{Lewis:2019xzd}. We use \texttt{MultiNest} interfaced with \texttt{MontePython} to sample the parameter space. The operational settings for \texttt{Multinest} that we use are shown in table \ref{tab:ns}\,\cite{Feroz:2007kg,Feroz:2008xx,Feroz:2013hea}.

   To constrain the model, we used the Planck 2018 likelihoods for the temperature and polarization power spectra\mycite{Aghanim:2019ame}. Here `TT+lowE' denotes the combination of low-$\ell$ TT $(\ell <30)$, low-$\ell$ EE and high-$\ell$ TT \texttt{plik-lite} $(\ell \geq 30)$ likelihood, and `TTTEEE+lowE' denotes the combination of low-$\ell$ TT, low-$\ell$ EE and high-$\ell$ TTTEEE \texttt{plik-lite} likelihood. In addition we also use Planck 2018 lensing likelihood. For BAO, we used the 6DF Galaxy survey, SDSS-DR7 MGS data, and the BOSS measurement of BAO scale and $f\sigma_8$ from DR12 galaxy sample\mycite{Beutler2011,Ross:2014qpa,Alam:2016hwk}\footnote{In this work, we used the latest corrected version of the BAO likelihood implemented in \texttt{MontePython3.3}. See Ref.\,\cite{bao} for more details.}. Finally for $H_0$ data, we used the latest measurement of local Hubble parameter from SH0ES collaboration\mycite{Riess:2019cxk}.
   We use the following combinations of likelihoods in our analysis: `TT+lowE', `TTTEEE+lowE', `TTTEEE+lowE+lensing', `TTTEEE+lowE+lensing+BAO', and `TTTEEE +lowE+lensing+BAO+$H_0$'. In table \ref{tab:prior}, we show the prior ranges used. We use a log-prior for the extra parameter $\geff$ because the expected features in its posterior distribution span over many orders of magnitude. Also, we do not vary the relativistic degrees of freedom $\neff$ in this analysis to disentangle the effect of only neutrino self-coupling on the CMB power spectra, fix it to $\neff = 3.046$.
   
   Furthermore, to analyse the MI and SI modes individually, we used separate priors for  $\lgeff$ to distinguish them. The prior ranges for the two modes for all datasets are shown in Table~\ref{tab:priorsep} below. The prior ranges are chosen such that the boundary between MI and SI modes is approximately at the minimum of the valley region between the two peaks in the posterior of $\lgeff$.
   \renewcommand{\arraystretch}{1.5}
   	\begin{table}[htb!]
   		\centering
   			\caption{Prior ranges on $\lgeff$ for the mode separation}
   		\begin{tabular}{l|M{3.5cm}|M{3.5cm}}
   			\hline\hline
   			Dataset & MI  & SI  \\
   			\hline
   			TT+lowE & $[-5.0,-2.3]$ & $[-2.3,-0.8]$ \\
   			\hline
   			All other dataset & $[-5.0,-2.7]$ & $[-2.7,-0.8]$ \\
   			\hline
   		\end{tabular}
   	\label{tab:priorsep}
   	\end{table}
  
	\renewcommand{\arraystretch}{1}
	\section{Results}\label{sec:result}
	\begin{table}[t!]
			\centering
			\caption{Parameter values and 68\% confidence limits in \three.}\label{tab:params3c}
			\begin{tabular}{c|M{2.68cm}M{2.68cm}|M{2.68cm}M{2.68cm}}
				\hline\hline
				Parameters & \multicolumn{2}{c}{TT+lowE}\vline & \multicolumn{2}{c}{TTTEEE+lowE} \\
				\hline
				& SI & MI & SI & MI\\\cline{2-5}
				$\omb$ & $0.022\pm 0.0003$ & $0.022\pm 0.00022$ & $0.022\pm 0.00016$ & $0.022\pm 0.00015$\\[1ex]
				$\omc$ & $0.1212\pm 0.0025$ & $0.1203\pm 0.0021$ & $0.1205\pm 0.0015$ & $0.1201\pm 0.0014$\\[1ex]
				$100\theta_s$ & $1.0469\pm 0.00068$ & $1.0419\pm 0.00048$ & $1.0464\pm 0.00087$ & $1.0419\pm 0.0003$\\[1ex]
				$\ln(10^{10}A_s)$ & $2.968\pm 0.0186$ & $3.036\pm 0.017$ & $2.984\pm 0.017$ & $3.042\pm 0.0161$\\[1ex]
				$n_s$ & $0.9317\pm 0.0085$ & $0.9593\pm 0.0071$ & $0.9386\pm 0.004$ & $0.9626\pm 0.005$\\[1ex]
				$\tau_\mathrm{reio}$ & $0.0501\pm 0.0082$ & $0.0516\pm 0.0079$ & $0.0543\pm 0.0077$ & $0.0537\pm 0.0077$\\[1ex]
				$\lgeff$ & $-1.72\pm 0.17$ & $-4.17\pm 0.51$ & $-1.92\pm 0.18$ & $-4.35\pm 0.42$\\[1ex]
				\hline
				$H_0(\hu)$ & $68.97\pm 1.05$ & $67.52\pm 0.93$ & $69.44\pm 0.64$ & $67.82\pm 0.61$\\[1ex]
				$r_s^* (\mathrm{Mpc})$ & $144.70\pm 0.53$ & $144.97\pm 0.49$ & $144.54\pm 0.35$ & $144.84\pm 0.32$\\[1ex]
				$\sigma_8$ & $0.826\pm 0.01$ & $0.824\pm 0.009$ & $0.834\pm 0.008$ & $0.824\pm 0.0075$\\[1ex]
				\hline
				$\chi^2 - \chi^2_{\Lambda\text{CDM}} $  & 2.33 & -0.01 & 5.14 & 0.18\\[1ex]
				\hline
			\end{tabular}
		\end{table}

\begin{figure}
	\centering
	\includegraphics[width=0.83\linewidth]{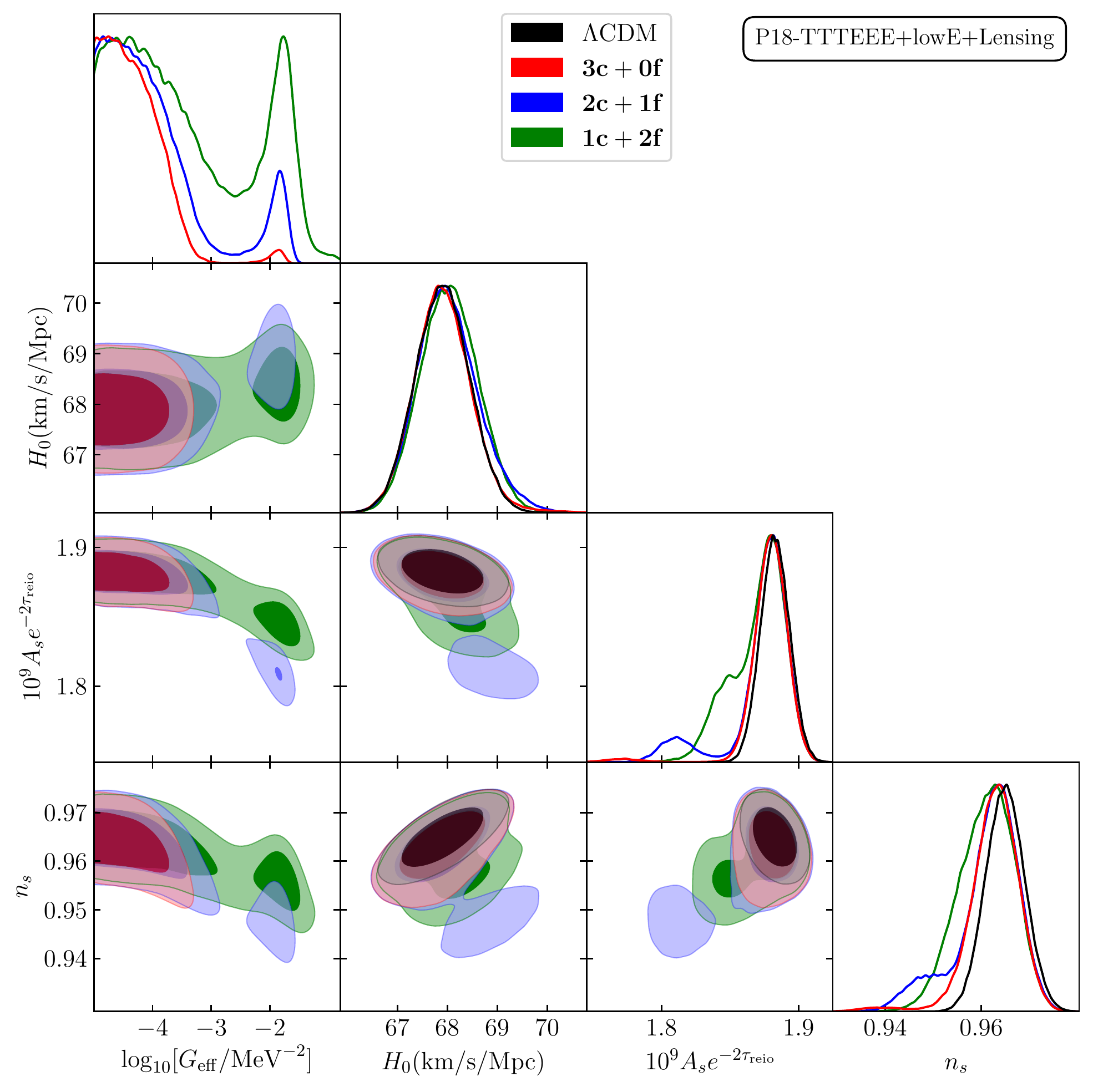}
	\caption{The $68\%$ and $95\%$ confidence limits of $\lgeff, H_0, A_se^{-2\tau_\mathrm{reio}}$, and $n_s$ for neutrino interaction scenarios \one{} (green) , \two{} (blue) and \three{} (red) for Planck 2018 TTTEEE+lowE+lensing dataset. For reference we also show the posteriors for \lcdm{} model (black) for the same dataset. We see that both $A_se^{-2\tau_\mathrm{reio}}$ and $n_s$ are negatively correlated with $\lgeff$, whereas $H_0$ is positively correlated.}
	\label{fig:p18ttteeelensing-3casewlc}
\end{figure}

	In this section, we present the results of our MCMC analysis. 
	In the \three{} scenario, we assume that all of three species self-interact and we use the same value of $\geff$ for all of them. This is because all of them are massless, and hence are equivalent. We have explicitly checked that even if we let the couplings for each species vary independently, their final posteriors are identical. Therefore, we use the same value of coupling without loss of generality. 
	Below we present the results in flavor universal and flavor-specific scenario. In figure~\ref{fig:p18ttteeelensing-3casewlc} we show the posteriors for $\lgeff$ and other relevant parameters for all three \sinu{} scenarios and \lcdm{}.

\subsection{Flavor-universal scenario: Three-coupled states (\three)}\label{sec:flavor-universal}
	This is the only scenario which has been analysed before in Refs.\mycite{2014PhRvD..90l3533C,Lancaster:2017ksf,Kreisch:2019yzn} albeit with 2015 and older Planck likelihoods. We show the results here using the Planck 2018 likelihood for the first time. The inferred parameter values and their 68\% confidence limits for the TT+lowE and TTTEEE+lowE datasets are given in table\,\ref{tab:params3c}.

\begin{figure}
		\begin{subfigure}{\linewidth}
			\centering
			\includegraphics[width=\linewidth]{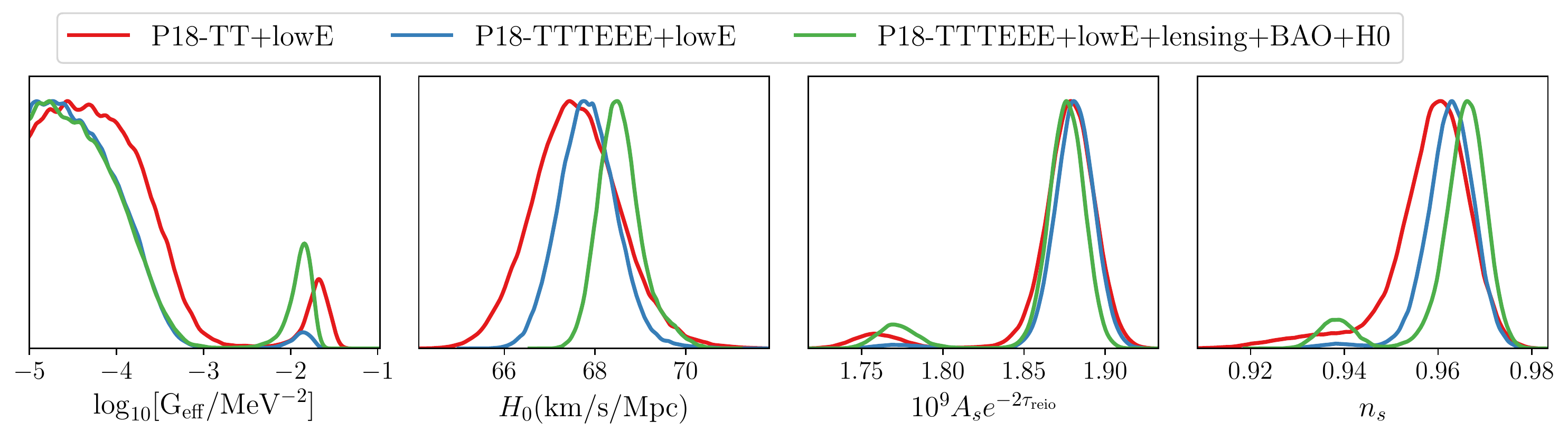}
			\subcaption{\three}\label{fig:3c-1d-pos}
		\end{subfigure}\vspace{0.45cm}
		\begin{subfigure}{\linewidth}
			\centering
			\includegraphics[width=\linewidth]{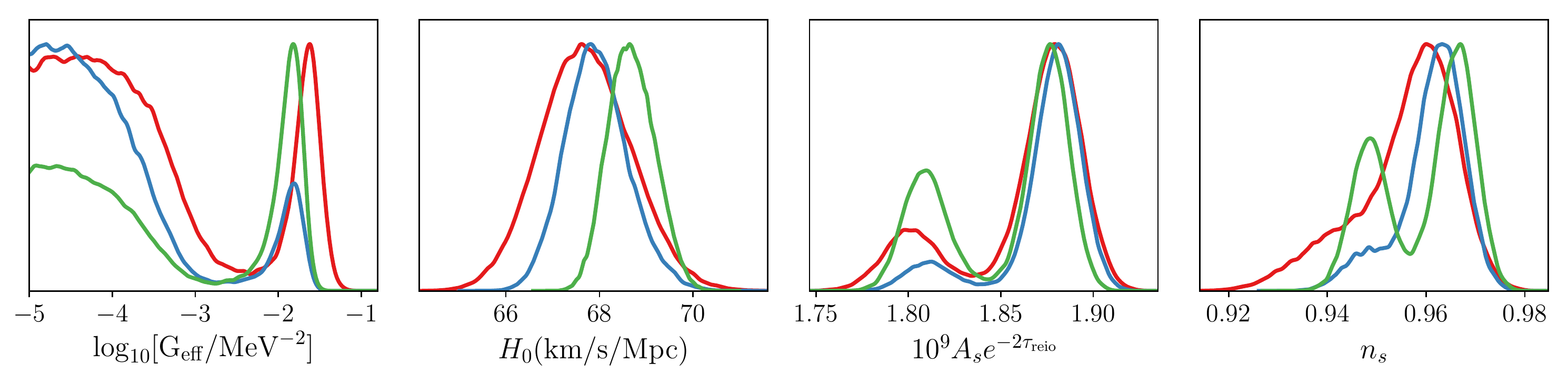}
			\subcaption{\two}\label{fig:2c-1d-pos}
		\end{subfigure}\vspace{0.45cm}
		\begin{subfigure}{\linewidth}
			\centering
			\includegraphics[width=\linewidth]{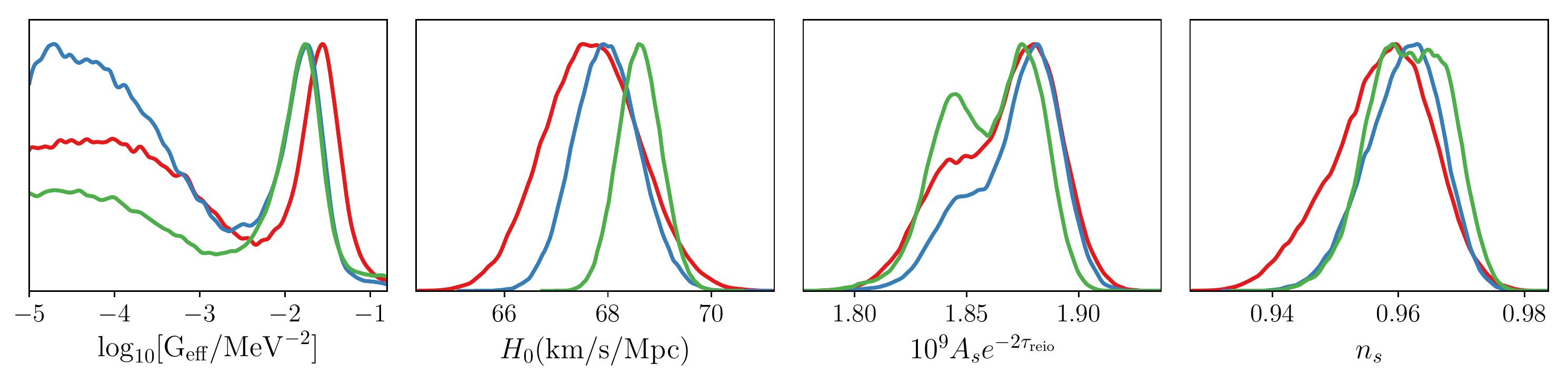}
			\subcaption{\one}\label{fig:1c-1d-pos}
		\end{subfigure}
		\caption{Marginalised 1D posteriors for $\lgeff, H_0, A_se^{-2\tau_\mathrm{reio}}$, and $n_s$ for three \sinu{} scenarios - \three{} (top), \two{} (middle) and \one{} (bottom) for three datasets.}\label{fig:1d-pos}
	\end{figure}

	Our results for the posteriors for $\lgeff$ and other relevant parameters are shown in figure~\ref{fig:3c-1d-pos}. 
	We find a multimodal posterior for $\lgeff$ in agreement with the previous analyses. The SI mode corresponds to a larger value of $\lgeff = -1.92\pm 0.18$, whereas the MI mode is characterized by $\lgeff = -4.35\pm 0.42$  for the TTTEEE+lowE dataset (see table\,\ref{tab:params3c}). These values do not change appreciably for other datasets, except for TT+lowE, the reason of which we discuss later. The most important effect of including the Planck 2018 polarization data is that it suppresses the significance of the SI mode substantially compared to TT+lowE. Inclusion of the polarization data also shifts the $\lgeff$ posterior towards smaller value as can be seen in the marginalized posteriors in figure\,\ref{fig:3c-1d-pos}. In the rest of this subsection, we explain different aspects of these results.
	
	\begin{figure}[t]
		\centering
		\includegraphics[width=1.03\linewidth]{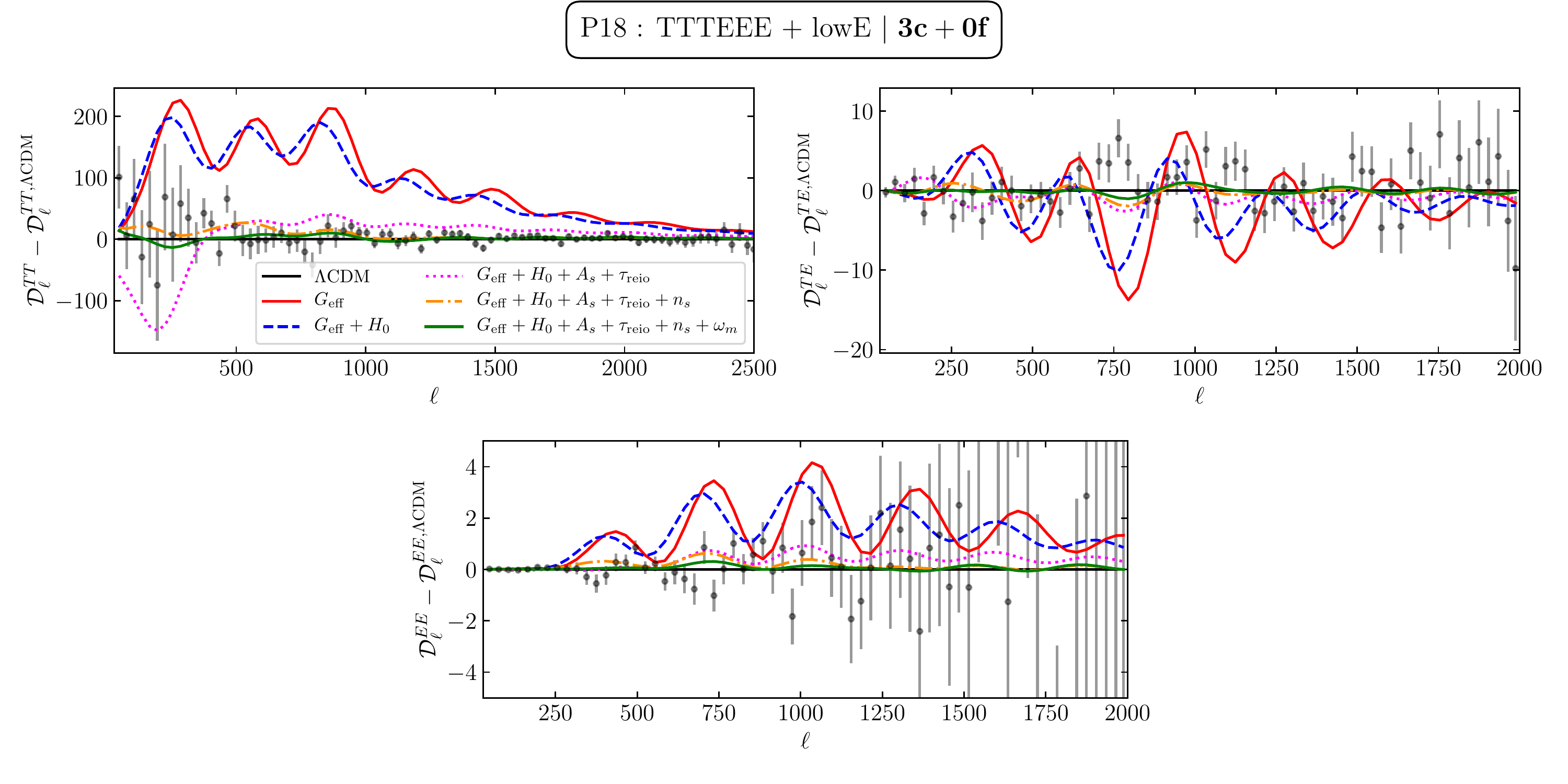}
		\caption{The residuals of the CMB TT (top left), TE (top right), and EE (bottom) power spectra starting from \lcdm{} best-fit, and with successive inclusion of the SI mode best-fit values of $\geff, H_0, A_s, \tau_\mathrm{reio}, n_s$, and $ \omega_m\equiv\Omega_{\rm b}h^2+\Omega_{\rm c}h^2 $, respectively, in \three. Starting from $\geff$, $H_0$ corrects for the phase of the spectrum, $A_s$ and $\tau_\mathrm{reio}$ reduce the amplitude, and $n_s$ red-tilts the whole spectrum. See section\,\ref{sec:flavor-universal} for more details.}
		\label{fig:TT_residuals}
	\end{figure}
	
	First, let us try to understand the origin of the SI mode which can be explained using the degeneracy of $\geff$ with parameters. To this end, we first show the changes in the residuals of the TT, TE, and EE power spectra in figure\,\ref{fig:TT_residuals} as we change $\geff$ and the other correlated parameters, successively, to their SI mode best-fit values for TTTEEE+lowE dataset, starting from the \lcdm{} best-fit spectrum. These other parameters are $H_0, A_s, \tau_\mathrm{reio}, n_s$, and $\omega_m$, respectively. 
	When we first incorporate $\geff$ (red, solid), we see that the spectrum moves upwards and shifts towards larger $\ell$ according to our discussion in the previous section. This gives rise to the positive residuals and the oscillations. Next, we change the best-fit value of $H_0$ (blue, dashed) which primarily compensates for the phase shift of the spectrum. What remains after this is mostly an overall amplitude offset  barring a small amount of residual phase-shift. The amplitude offset is taken care of by the best-fit values of $A_s$ and $\tau_\mathrm{reio}$ (magenta, dotted). A smaller value of $A_s$ and a larger $\tau_\mathrm{reio}$ suppress the amplitude bringing down the residuals very close to the \lcdm{} values. This overall amplitude change over-compensates the modifications at very large scale which entered the horizon after neutrino decoupling, resulting in the dip in the $\ell\sim 200$ region. A smaller value of the $n_s$ (orange, dot-dashed) corrects for this  and red-tilts the spectrum increasing power at low $\ell$ while suppressing power at large $\ell$. Finally, a smaller $\omega_{\rm m}$ (green, solid) reduces the residuals even more, bringing it very close to the \lcdm{} spectrum. A similar behavior is shown by the EE spectrum as well which, however, has much larger error bars at high $\ell$ than TT, and has less constraining power. This exercise shows that the Planck CMB data allows for a larger value of $\geff$ using its degeneracy with other parameters giving rise to the SI mode. However, this compensation mechanism works for a special range of $\geff$ values where it is large enough to enhance all the acoustic peaks so that the effects can be compensated by other global parameters, and also the same time does not impart very large modification of the first acoustic peak and large phase shift. \cb{For very high values of $\geff$, neutrino remains strongly coupled till very late times, even after recombination, and behaves almost always like a perfect fluid before recombination, and this scenario is disfavored by CMB data~\cite{Baumann:2015rya}}. All of these prefer $\geff$ values for which neutrino decoupling happens slightly prior matter-radiation equality in the SI mode.
	\cb{The origin of the MI mode is rather easier to understand. For smaller value of $ \geff $, the changes in the spectrum are very small compared to the \lcdm{}. This results in the plateau at small value in the posterior of $\lgeff$. }

\begin{figure}[t!]
	\centering
	\begin{subfigure}{0.49\linewidth}
		\includegraphics[width=\linewidth]{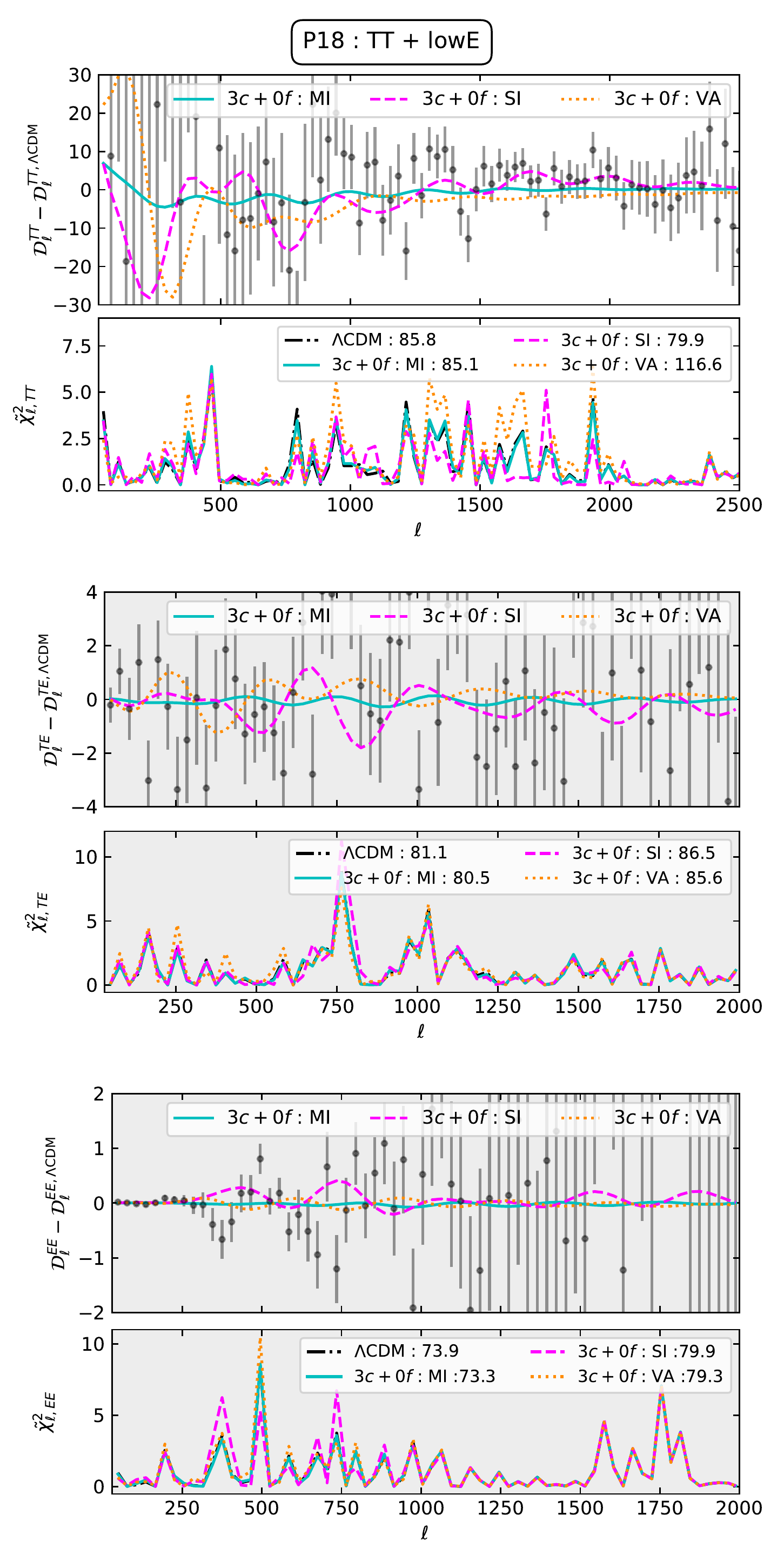}
	\end{subfigure}
	\begin{subfigure}{0.49\linewidth}
	\includegraphics[width=\linewidth]{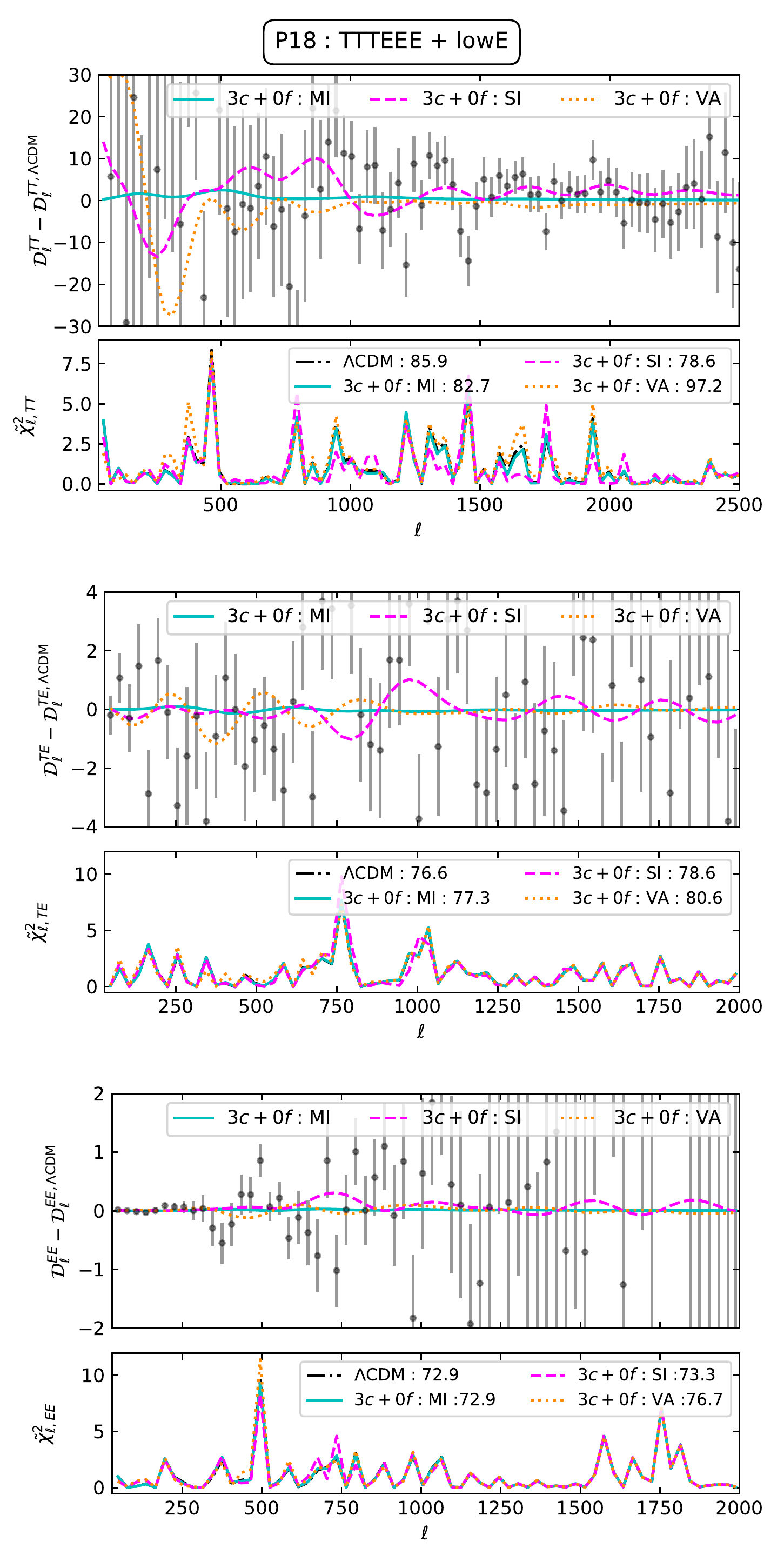}
    \end{subfigure}
\caption{Residual plots relative to the \lcdm{} best-fit for high-$ \ell $ ($ \ell = 29-2500 $ for TT and $ \ell = 29-2000 $ for TE and EE) modes for \sinu{} with TT+lowE (left) and TTTEEE+lowE(right) datasets. In the sub-panel for each residual plot, we show the distribution of $ \tilde{\chi}_\ell^2 $ ( see Eq.\eqref{eq:chisq_ell}) with $ \ell $ for \lcdm{} and SINU modes. In the legend of each plot, we show the total $ \tilde{\chi}^2 $. The gray-shaded TE (middle-left)and EE (bottom-left) plots for TT+lowE signifies that high-$ \ell $ TE and EE mode data are not included in that analysis. VA denotes the `best-fit' point in the `valley' between the two modes in $\geff$ posterior.}
\label{fig:chisq_ell}
\end{figure}

    Now we shall give a quantitative explanation for the significance of the SI mode relative to the MI mode. From figure\,\ref{fig:3c-1d-pos}, clearly the significance of the former depends on the dataset used. To better understand the origin of this variation in significance, we first define a quantity \achil{} as
    \begin{equation}\label{eq:chisq_ell}
    \text{\achil} \equiv \frac{(\mD_\ell^{\rm BF} - \mD_\ell^{\rm Planck})^2}{\sigma_\ell^2}\,,
    \end{equation}
    where $D_\ell^{\rm Planck}$ is the binned Planck data, and $D_\ell^{\rm BF}$ is the power spectrum corresponding to our best-fit point, and $\sigma_\ell$ is the error bar of the Planck binned data. Therefore, \achil{} carries information about  the \emph{goodness-of-fit} of the spectrum in different regions of $\ell$, and gives an idea about which part of the spectrum prefers/penalizes the fit. We define a quantity \achi{} as $\text{\achi} = \sum_\ell \text{\achil}$ which gives the approximate \emph{goodness-of-fit} information in the whole spectrum. Here we are using the binned Planck data for visual clarity in the plots. In \achi{}, we ignore the bin-by-bin correlation of the CMB data.  Note that, the \achi{} is not the same quantity as the $\chi^2$ calculated using the full Planck likelihood.
    

    In figure \ref{fig:chisq_ell}, we plot the residuals and \achil{} for TT, TE, and EE spectrum in the high-$\ell$ region for the SI and MI mode best-fit points in \three{} for the TT+lowE (left panels) and TTTEEE+lowE (right panels) datasets\footnote{In the low-$\ell$ region, the residuals are small compared to the errorbar in the data.}. We also show the corresponding plots for the best-fit point in the `valley' (VA) between the two modes which is defined within the range $-2.6 < \lgeff < -2.4$.\footnote{The best-fit point in the valley region is the sampling point that has the smallest $\chi^2$ in that range.} We define the VA point to help explain the origin behind the \emph{separate} SI mode. Firstly, we note that the MI residuals are very small and very close to the \lcdm{} points, i.e., the x-axis. This is expected as the CMB data loses any sensitivity for $\geff$ values below the MI mode limit, and the spectrum becomes practically indistinguishable from \lcdm. Interestingly however, the SI mode best-fit point yields a better fit to the \lcdm{} residuals which is evident in the TT spectrum in the top panels of figure\,\ref{fig:chisq_ell}. This substantially enhances the significance of the SI mode which is reflected in the \achil{} distribution. In several $\ell$-regions, e.g. around $\ell \simeq 750, 1300, 1700$, and $1900$, the SI mode \achil{} is below the \lcdm{} or the MI mode values, which largely compensates for the other $\ell$ regions with relatively worse fit. These are exactly the places where the SI best-fit point fits the \lcdm{} residuals. In fact, according to the simplified analysis, the SI mode \achi{} for the TT+lowE data is better than both MI mode or \lcdm{}. 
    
    This is, however, not true for the TE and EE spectra though. The peaks in the EE spectrum are sharper than in the TT spectrum. Therefore, the polarization spectrum is more sensitive to phase shift due to \sinu{} than the temperature spectrum, even though the error bars are larger\,\cite{Baumann:2015rya}. This is reflected in the bottom panels in figure\,\ref{fig:chisq_ell}, where we see that the SI mode \achil{} is always equal to or above the MI or \lcdm{} plots, yielding a poorer fit especially to the low and intermediate-$\ell$ polarization data. Similar behaviour is also observed in the TE residuals (middle panels). As a result, when the polarization data is included, the significance of the SI mode is reduced.
    
    The presence of the `valley' between the two modes can be understood from the lowest $\ell$ mode that is affected by $\geff$. From the top axis of figure\,\ref{fig:opacity}, we see that a typical SI mode $\geff$ affects all modes $\ell \gtrsim 200$ which happens to be the approximate position of the first peak in the CMB spectrum. Therefore, it is easier to compensate for the modifications by changing the other parameters, as explained earlier, yielding a reasonably good fit. On the other hand, MI mode values of $\geff$ affects only $\ell \gtrsim 10^4$ which are far beyond the range of the present CMB experiments. Planck measured the TT spectrum only upto $\ell=2500$, and the EE spectrum upto $\ell=2000$. Therefore, the MI mode is virtually indistinguishable from \lcdm{} as far as the Planck data is concerned. For intermediate values of $\geff$ between the two modes, the CMB spectrum is modified only in the high-$\ell$ part of the Planck $\ell$-range. In this case the degeneracy with other parameters, which impart changes in whole spectrum, cannot be exploited to get an overall good fit. Thus, a `valley' appears between the two modes as a result of the poor fit to the CMB data.
    
	The polarization data (TTTEEE+lowE) also shifts of the whole posterior $\geff$ to the left relative to TT+lowE which evident from the left panels of figure\,\ref{fig:1d-pos}. This can be understood from the relative sizes of the errorbars in the TT, TE, and EE spectra. The TT spectrum has smaller errorbars at large $\ell$. Whereas, the TE and EE spectra uncertainties are smaller at small $\ell$ (see figure\,\ref{fig:chisq_ell}). This essentially means that the TT data can accommodate large deviations at small-$\ell$, whereas TE and EE data have more freedom at larger-$\ell$. Therefore, inclusion of the polarization data penalizes any deviation at low $\ell$. Because, stronger self-interaction implies a later decoupling of the neutrinos affecting relatively smaller $\ell$, TTTEEE+lowE data prefers a slightly smaller value of $\geff$ compared to TT+lowE.

	\begin{figure}[t]
	\centering
	\includegraphics[width=\linewidth]{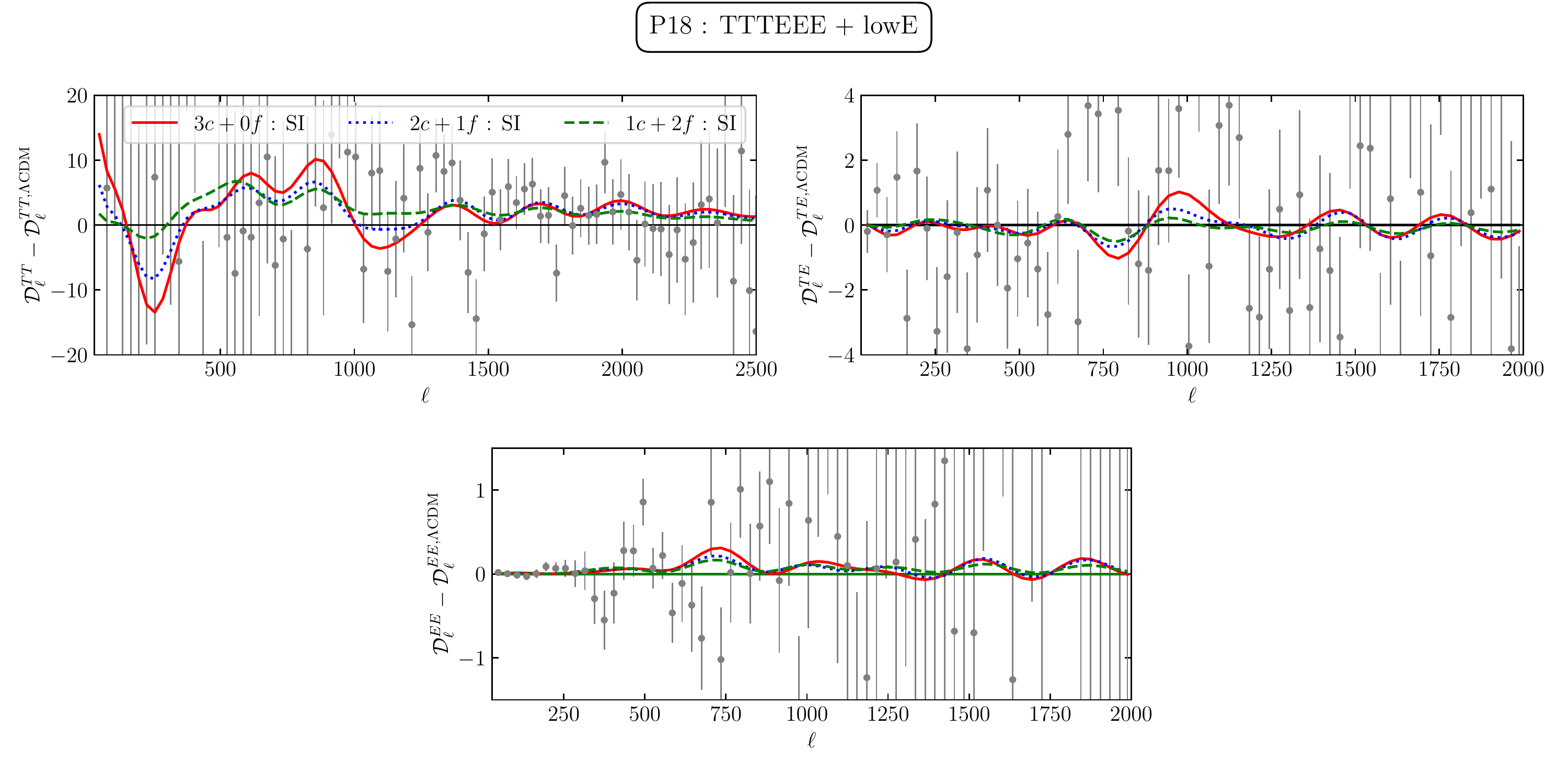}
	\caption{The residuals of the SI mode best-fit spectra of TT (top left), TE (top right), and EE (bottom) relative to \lcdm{} in three scenarios. We note that these curves exhibit a better fit to the \lcdm{} residuals which shown in gray. circles.}\label{fig:residual_3_cases}
	\end{figure}

	\subsection{Flavor-specific scenario: Two-coupled (\two) \& one-coupled states (\one)}\label{sec:flavor-specific}
	In this section, we shall concentrate on the flavor-specific neutrino self-interaction scenarios, namely, \two{} with only two self-interacting neutrino states, and \one{} with only one self-interacting state. For the results in this section, we used the same coupling strength for the two interacting states instead of having two independent couplings in the MCMC sampling. We have explicitly verified the validity of this assumption by running chains with independent couplings and same prior for the states which yielded same posterior distributions. This was rather expected as all states are massless, and are not distinguishable from one another as far as their cosmology is concerned. Hence, we chose the same coupling to derive all the results here reducing the number of free parameters without loss of generality. In passing, we also note that the one could have chosen different priors for different states inspired by the bounds from BBN and laboratory experiments which would yield distinct posteriors\,\cite{Blinov:2019gcj,Lyu:2020lps}. We want to point out that \one{} scenario is equivalent to the flavor-specific $\nu_e/\nu_\mu/\nu_\tau$ coupling cases discussed in Ref.~\cite{Blinov:2019gcj}.
	
	The posteriors for $\lgeff$ are shown in figure\,\ref{fig:p18ttteeelensing-3casewlc}, and the inferred parameter values and their 68\% confidence limits for the TT+lowE and TTTEEE+lowE datasets are given in table\,\ref{tab:params2c} for \two. The two modes are present in both of these cases. The SI mode appears at almost at the same value of $\geff$ as in \three. As discussed before, this specific value of $\geff$ is determined by the $\ell$-range of the Planck data. The interplay between $\geff$ and other degenerate parameters seeks out this value. At the same time there is a very small shift in the SI mode $\geff$ values across three \sinu{} scenarios. However, the most striking difference between \three{} and \two{} or \one{} is the \emph{enhancement} of the SI modes for all datasets in the latter, as can be seen from figure~\ref{fig:1d_posterior} and \ref{fig:p18ttteeelensing-3casewlc} (see also figure\,\ref{fig:2c-tri-pos} and \ref{fig:1c-tri-pos}). The enhancement is more pronounced in \one{} than \two{}.  From Using Eq.~(\ref{eq:phase_shift}), (\ref{eq:amp_supp}), \eqref{eq:rnuval} and we see that
	 the phase shift and the amplitude suppression due to neutrino free-streaming are proportional to the number of free-streaming neutrino states. With less number of self-interacting states, these changes in the CMB spectra are relatively milder compared to \three{}. This is evident from figure~\ref{fig:cmb_spectra}.  As a result, there exists more room to use the degeneracy between $\geff$ and other parameters, especially $A_s$ and $n_s$, to achieve a better fit. This is seen in figure\,\ref{fig:residual_3_cases} where we see that SI mode the residuals in the flavor-specific scenario are always smaller than the flavor universal case. 
	
	Another new feature here is that the upward rise of the valley between the two modes in the $\lgeff$ posterior. In the previous section, we explained the origin of the valley as a result of the fact that intermediate values of $\geff$ affect only a part of the CMB spectra observed by Planck, which cannot be compensated by varying other degenerate parameters. Hence, the intermediate values are disfavored. However in \two{} or \one, those changes are comparatively modest, and can be partly undone by varying other parameters. As a result, the valley points yield smaller $\chi^2$ compared to \three.

	\begin{table}[t]
		\centering
		\caption{Parameter values and 68\% confidence limits in \two.}\label{tab:params2c}
		\begin{tabular}{c|M{2.68cm}M{2.68cm}|M{2.68cm}M{2.68cm}}
			\hline\hline
			Parameters & \multicolumn{2}{c}{TT+lowE}\vline & \multicolumn{2}{c}{TTTEEE+lowE} \\
			\hline
			& SI & MI & SI & MI\\\cline{2-5}
			$\omb$ & $0.022\pm 0.00027$ & $0.022\pm 0.00021$ & $0.022\pm 0.00016$ & $0.022\pm 0.00015$\\[1ex]
			$\omc$ & $0.1211\pm 0.0023$ & $0.1203\pm 0.002$ & $0.1205\pm 0.0014$ & $0.1201\pm 0.0013$\\[1ex]
			$100\theta_s$ & $1.0452\pm 0.00059$ & $1.0419\pm 0.0005$ & $1.045\pm 0.00076$ & $1.0419\pm 0.00031$\\[1ex]
			$\ln(10^{10}A_s)$ & $2.99\pm 0.0179$ & $3.036\pm 0.01714$ & $3\pm 0.0167$ & $3.042\pm 0.0161$\\[1ex]
			$n_s$ & $0.9407\pm 0.0079$ & $0.9596\pm 0.0068$ & $0.9473\pm 0.0046$ & $0.9628\pm 0.005$\\[1ex]
			$\tau_\mathrm{reio}$ & $0.0501\pm 0.008$ & $0.0516\pm 0.0079$ & $0.0538\pm 0.0077$ & $0.0538\pm 0.0077$\\[1ex]
			$\lgeff$ & $-1.69\pm 0.2$ & $-4.03\pm 0.6$ & $-1.93\pm 0.24$ & $-4.24\pm 0.5$\\[1ex]
			\hline
			$H_0(\hu)$ & $68.34\pm 1.00$ & $67.57\pm 0.92$ & $68.81\pm 0.63$ & $67.83\pm 0.6$\\[1ex]
			$r_s^* (\mathrm{Mpc})$ & $144.75\pm 0.51$ & $144.98\pm 0.49$ & $144.64\pm 0.34$ & $144.85\pm 0.32$\\[1ex]
			$\sigma_8$ & $0.823\pm 0.01$ & $0.824\pm 0.009$ & $0.829\pm 0.0079$ & $0.824\pm 0.0075$\\[1ex]
			\hline
			$\chi^2 - \chi^2_{\Lambda\text{CDM}}$ & -0.17 & -0.05 & 1.8 & 0.28\\[1ex]
			\hline
		\end{tabular}
	\end{table}
    \begin{table}[t]
    	\centering
    	\caption{Parameter values and 68\% confidence limits in \one.}\label{tab:params1c}
    	\begin{tabular}{c|M{2.68cm}M{2.68cm}|M{2.68cm}M{2.68cm}}
    		\hline\hline
    		Parameters & \multicolumn{2}{c}{TT+lowE}\vline & \multicolumn{2}{c}{TTTEEE+lowE} \\
    		\hline
    		& SI & MI & SI & MI\\\cline{2-5}
    		$\omb$ & $0.022\pm 0.00023$ & $0.022\pm 0.00021$ & $0.022\pm 0.00015$ & $0.022\pm 0.00015$\\[1ex]
    		$\omc$ & $0.1207\pm 0.0021$ & $0.1203\pm 0.002$ & $0.1203\pm 0.0014$ & $0.1201\pm 0.0013$\\[1ex]
    		$100\theta_s$ & $1.0434\pm 0.00062$ & $1.0419\pm 0.0004$ & $1.043\pm 0.00058$ & $1.0419\pm 0.0003$\\[1ex]
    		$\ln(10^{10}A_s)$ & $3.01\pm 0.0179$ & $3.037\pm 0.01664$ & $3.024\pm 0.0166$ & $3.042\pm 0.016$\\[1ex]
    		$n_s$ & $0.9513\pm 0.0069$ & $0.9609\pm 0.0059$ & $0.9553\pm 0.0049$ & $0.963\pm 0.005$\\[1ex]
    		$\tau_\mathrm{reio}$ & $0.051\pm 0.008$ & $0.0519\pm 0.008$ & $0.0539\pm 0.0076$ & $0.0539\pm 0.0077$\\[1ex]
    		$\lgeff$ & $-1.75\pm 0.4$ & $-3.94\pm 0.6$ & $-1.9\pm 0.37$ & $-4.06\pm 0.6$\\[1ex]
    		\hline
    		$H_0(\hu)$ & $67.9\pm 1.00$ & $67.56\pm 0.93$ & $68.3\pm 0.62$ & $67.83\pm 0.61$\\[1ex]
    		$r_s^* (\mathrm{Mpc})$ & $144.88\pm 0.5$ & $144.96\pm 0.5$ & $144.76\pm 0.32$ & $144.84\pm 0.31$\\[1ex]
    		$\sigma_8$ & $0.821\pm 0.01$ & $0.823\pm 0.009$ & $0.825\pm 0.0083$ & $0.824\pm 0.0075$\\[1ex]
    		\hline
    		$\chi^2 - \chi^2_{\Lambda\text{CDM}}$ & -0.91 & -0.03 & 0 & 0.1 \\[1ex]
    		\hline
    	\end{tabular}
    \end{table}
	
    
    \subsection{Effects on $H_0$ and $\sigma_8$}
\begin{figure}[t]
    	\centering
    	\includegraphics[width=0.5\linewidth]{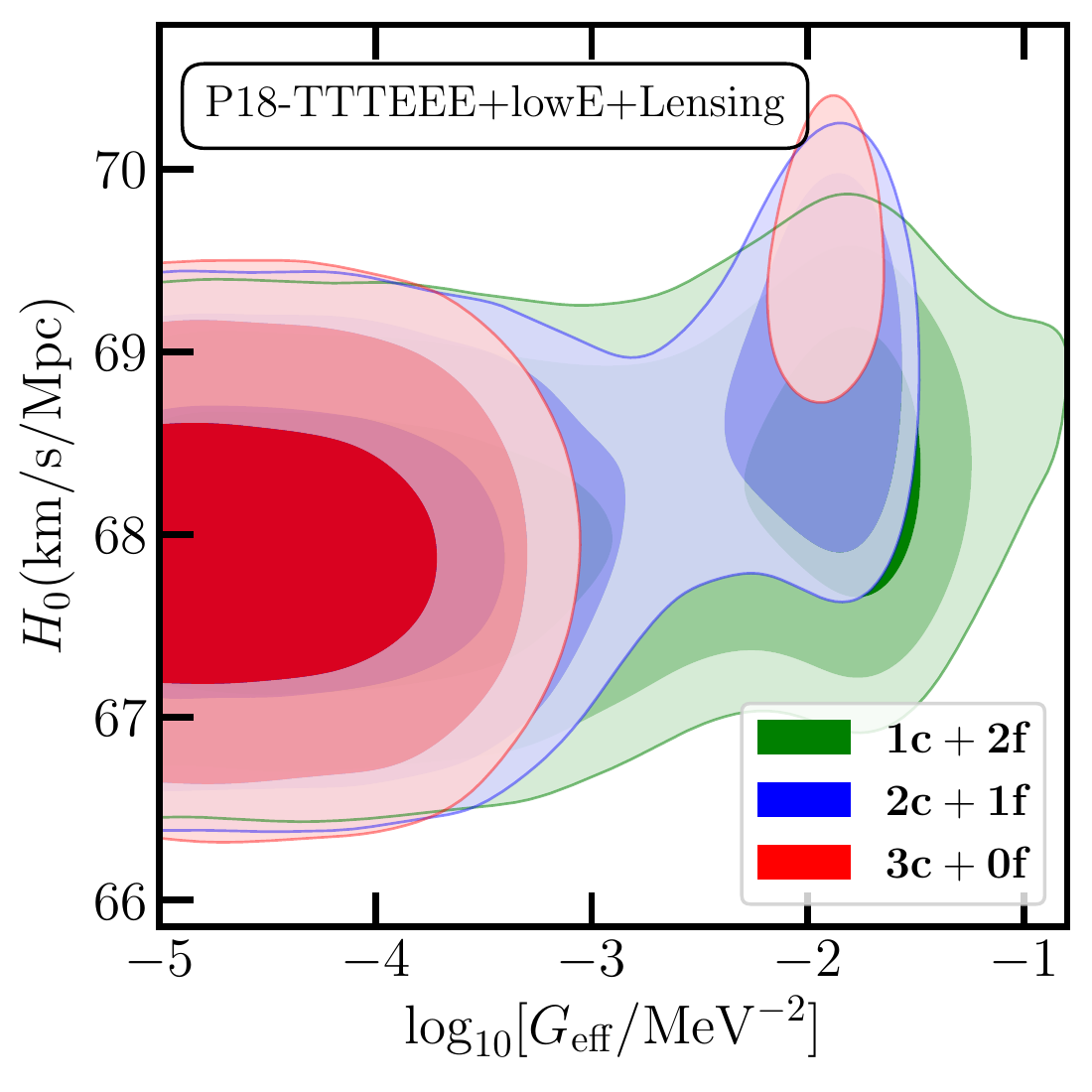}
    	\caption{The contours of 68\%, 95\%, and 99\% confidence levels for $\lgeff$ and $H_0$ in \one{} (green), \two{} (blue), and \three{} (red) for the Planck TTTEEE+lowE+lensing dataset. The SI mode contour for \three{} yields the largest value of $H_0$ because of bigger phase shift. However, the significance of the mode is much less compared to \two{} and \one{}.}
    	\label{fig:2d-h0}
    \end{figure}

    We know that neutrino self-coupling strength $\geff$ has a positive correlation with $H_0$ through the phase-shift as explained in section\,\ref{sec:changes_in_CMB}. From table\,\ref{tab:params3c}, we see that the mean value of the Hubble shifts to $H_0=69.46\pm0.52\hu$ for the TTTEEE+lowE dataset for the SI mode, reducing the tension with the SH0ES measurement to a modest $\sim3\sigma$. On the other hand, when compared with the CCHP measurement $H_0=69.8\pm 1.9\hu$ which uses the tip of the red giant branch calibration, our value of $H_0$ is fully consistent with that\,\cite{freedman2019}. However, the SI mode value of $H_0$ slightly decreases in the flavor-specific \two{} and \one{} scenarios because of smaller phase shift, but the significance of the mode is increased. This is evident from figure~\ref{fig:2d-h0}. In table~\ref{tab:h0}, we show $H_0$, $\Omega_\Lambda$, $100\theta_s$, $r_s^\ast$, and $D_A^\ast$ for \three{}, \two{} and \lcdm{}. The shift in $\theta_s$ due to the phase shift in \sinu{} decreases $D_A^\ast$ that helps increase the value of $H_0$ and $\Omega_\Lambda$. Note that, the value of $r_s^\ast$ changes by  only $\sim 1\sigma$ from \lcdm{}, and plays a sub-dominant role in changing the Hubble constant.
    
   	\begin{table}[b]
	\centering
	\caption{Parameter values and 68\% confidence limits for SI mode in \three{} and \two, and \lcdm{} in TTTEEE+lowE+lensing data.}\label{tab:h0}
	\begin{tabular}{c|M{3cm}|M{3cm}|M{3cm}}
		\hline
		\hline
		& SI: \three& SI: \two & \lcdm{}\\
		\hline
		$ H_0 (\hu)$ &$ 69.47\pm 0.59 $ &$ 68.87\pm0.58 $ &$67.90\pm0.54$\\[1ex]
		$\Omega_\Lambda$&$ 0.7035\pm0.0071 $&$ 0.6989\pm0.0072 $&$ 0.6912 \pm 0.0073 $\\[1ex]
		$ 100\theta_s $&$ 1.0463\pm 0.00094$&$ 1.0447 \pm 0.00079 $&$1.04186\pm 0.00029 $\\[1ex]
		$r_s^\ast({\rm Mpc})$&$ 144.58\pm0.32 $&$ 144.69 \pm 0.31 $&$ 144.87\pm0.29 $\\[1ex]
		$D_A^\ast({\rm Mpc})$&$ 12.69\pm 0.036 $&$ 12.72 \pm 0.034$&$ 12.773\pm 0.028$\\[1ex]
		\hline
	\end{tabular}
\end{table}
    
    
    The significance of the SI mode is enhanced when the local Hubble measurement data $H_0=74.0 \pm 1.4\,{\rm km\,s^{-1}Mpc^{-1}}$ from SH0ES is included\,\cite{riess2016}. At the same time, it suppresses the MI mode significantly as can be seen from figure~\ref{fig:1d-pos}. This is not surprising given the fact that the SH0ES likelihood includes only Hubble data, and favors SI mode which yields a larger $H_0$. However, one should be cautious about combining the Planck with the SH0ES data because of the more than $\sim 4\sigma$ discrepancy between the two. Even though we show the results with Planck+lensing+BAO+SH0ES here, we do not draw any conclusion from them. Note that, the \three{} scenario yields higher $H_0$ but its significance is smaller. In \two{} and \one{} scenarios, although the significance of the SI mode is increased, the corresponding value of $H_0$  decreases due to smaller phase shift. Therefore, both flavor-universal and flavor-specific \sinu{} scenarios fail to yield a large enough value of $H_0$ to completely solve the Hubble tension (see figure~\ref{fig:1d-pos}). The parameters inferred from the combined TTTEEE+lowE+lensing+BAO+H0 dataset are quoted in table~\ref{tab:params3c_app} and \ref{tab:params2c_app}. In appendix\,\ref{sec:app3}, we show the results for the model \one$+\dNeff$, i.e., when one neutrino species is self-interacting and additional free-streaming radiation with effective neutrino number $\dNeff$ is present. The value of $ H_0 $ increases slightly compared to the \one{} scenario due to presence of the additional radiation relics which can be seen from Table~\ref{tab:1c2fNeff}.

%
    Apart from $H_0$, another parameter that shows a small discrepancy between CMB and low redshift data is the late-time matter clustering amplitude $\sigma_8$. Planck 2018 results found \mbox{$\sigma_8 = 0.811 \pm 0.006$} using TTTEEE+lowE+lensing data\,\cite{Aghanim:2018eyx}. However, the gravitational weak lensing observation by KiDS-1000 found $\sigma_8 = 0.76^{+0.025}_{-0.020}$\,\cite{2020arXiv200715632H}. There is about $\sim 2.4\sigma$ discrepancy between the two values. The value of $\sigma_8$ in \sinu{} cosmology is mainly determined by two competing effects. The lack of anisotropic stress in the neutrino bath boosts the perturbation modes which enter horizon before neutrino decoupling. In case of SI mode values of $\geff$, these modes happen to be in range of $\sigma_8$, and thereby boosting its value. On the other hand, the lower value of best-fit $A_s$ and $n_s$ suppress $\sigma_8$. As a result, $\geff$ and $\sigma_8$ show a very weak correlation as can be seen in figure\,\ref{fig:3c-tri-pos}, \ref{fig:2c-tri-pos}, and \ref{fig:1c-tri-pos}. We find $\sigma_8 = 0.834\pm 0.0067$ and $0.829\pm 0.0058$ in SI mode for \three{} and \two, respectively for TTTEEE+lowE (see table~\ref{tab:params3c} and \ref{tab:params2c}).
    
	\subsection{Implications of BAO measurements on \sinu}
	\begin{figure}[t]
		\centering
		\includegraphics[width=0.9\linewidth]{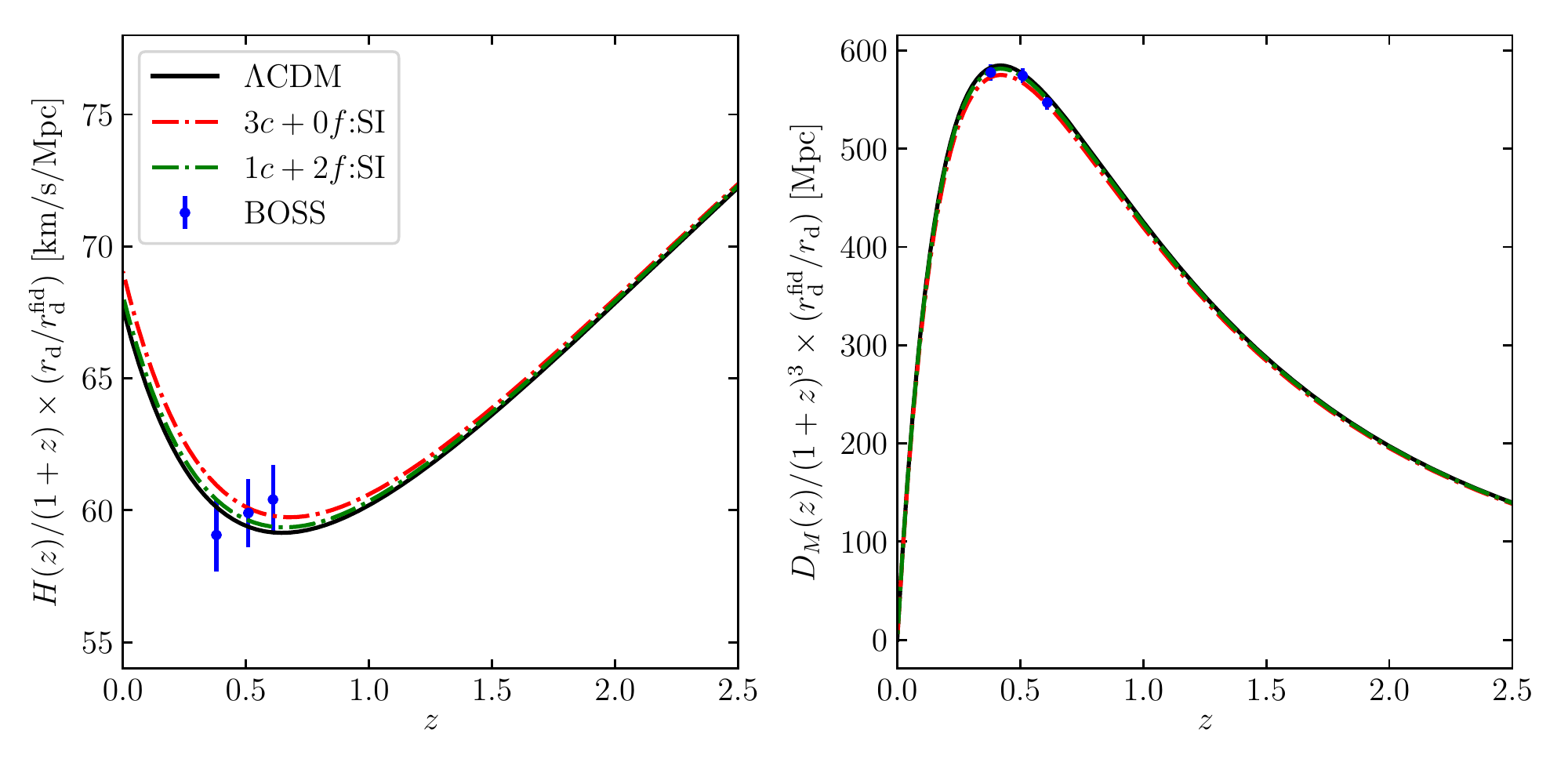}
		\caption{Evolution of $ H(z)/(1+z)\times (r_{\rm d} / r_{\rm d}^{\rm fid}) $ and $D_M(z)/(1+z)^3 \times ( r_{\rm d}^{\rm fid}/r_{\rm d} )$ bestfit \lcdm, and the SI modes of \three{} and \one{} for TTTEEE+lowE+lensing dataset. The different $ (1+z) $ scaling in both the plots are introduced for improved visualization of the data and the evolution graphs. The BOSS data points are taken from Ref.~\cite{alam2017}.}
		\label{fig:BAO}
	\end{figure}
	In this subsection, we examine in details the effects of the BAO measurements on \sinu{} cosmology. The BAO likelihood constraints the distance combinations $ D_M(z)/r_d $ and $ H(z)r_d $ where
	\begin{equation}\label{eq:rddef}
		r_d \equiv \int^\infty_{z_d} \dfrac{c_s(z)}{H(z)} \quad {\rm and } \quad D_M(z) \equiv (1+z)D_A(z)\,.
	\end{equation}
	Here, $r_d$ is the sound horizon at baryon-photon decoupling, $ z_d  $ is the redshift of that epoch, and and $ D_A(z) $ is the angular diameter distance\footnote{Note that, the phase shift in the CMB acoustic peaks due to interacting neutrinos will also induce phase shift in the BAO peaks\,\cite{Baumann:2017lmt,Baumann:2018qnt} (also see\,\cite{Ghosh:2019tab}). For SI mode in particular, the effect of this scale dependent phase shift of BAO peaks with respect to $ \Lambda $CDM cosmology can be important.
	The BAO scales extracted from the observed spectrum assuming fiducial $ \Lambda $CDM cosmology~\cite{Alam:2016hwk}, may differ from the true BAO scales in SINU cosmology due to the phase shift (see Ref.~\cite{Bernal:2020vbb} for quantitative study of the neutrino phase-shift on BAO scales in different cosmologies). In this work we ignore this subtlety and use the BAO likelihood as reported in Ref.~\cite{Alam:2016hwk}}. By the virtue of phase shift due to the strongly interacting neutrinos, the SI mode accommodates a larger Hubble constant compared to the MI mode, as discussed previously.

\begin{figure}[t!]
	\centering
	\includegraphics[width=0.6\linewidth]{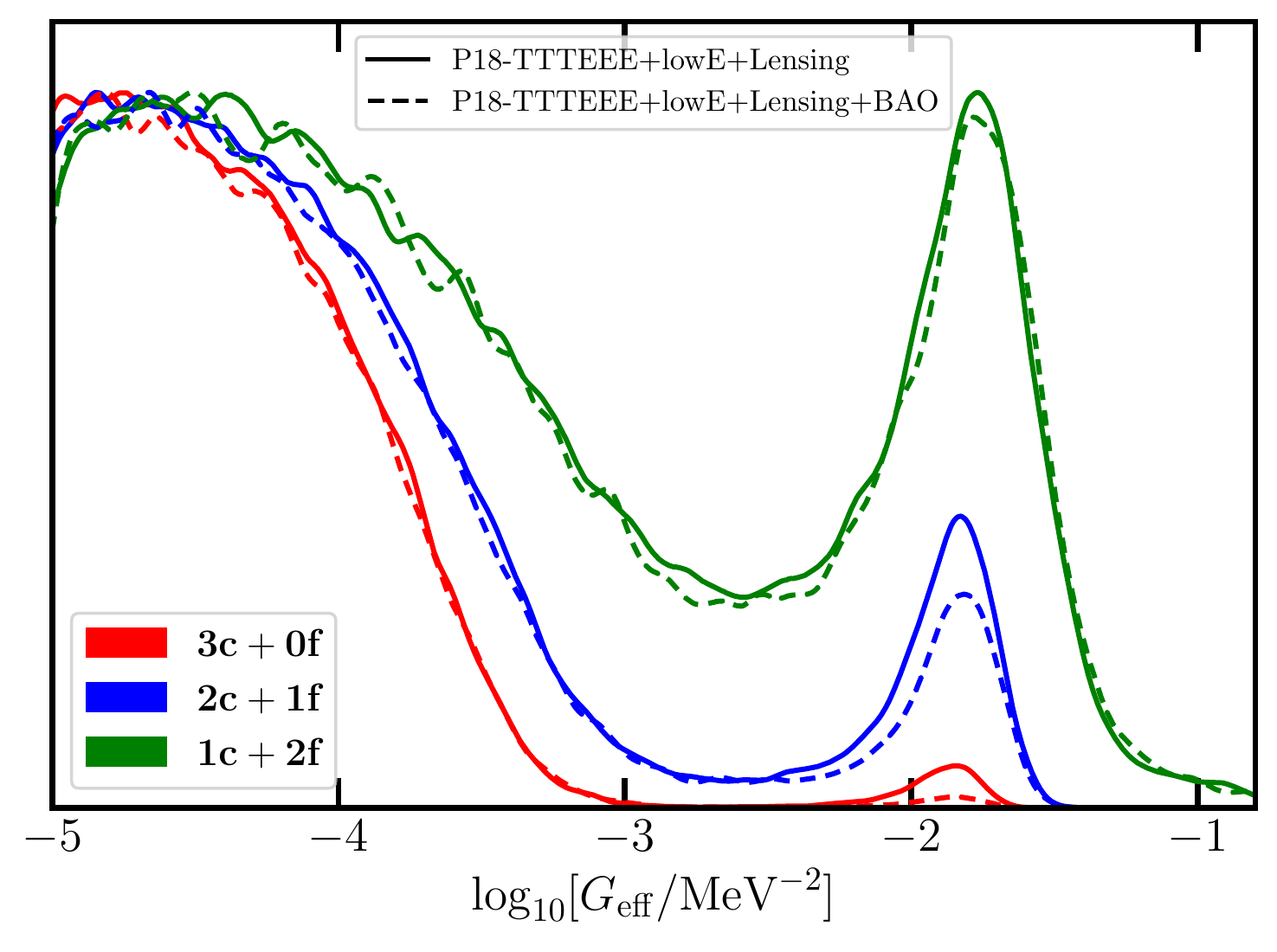}
	\caption{1-D posterior of $\lgeff$ for \one, \two, \three{} with TTTEEE+lowE+lensing and TTTEEE+lowE+lensing+BAO datasets. Inclusion of the BAO data reduces the relative signification of the SI mode peak compared to the MI mode. }
	\label{fig:baopos}
\end{figure}

We show in figure~\ref{fig:BAO} the evolution of $ H(z)r_d $ and $ D_M(z)/r_d $ (with additional scaling of the scale factor $ a = 1/(1+z) $) for the TTTEEE+lowE+lensing bestfit values in \three{} and \one{} scenarios along with the BOSS data for the BAO scales. It is evident from the figure that the BOSS data mildly disfavors the SI mode compared to the MI mode (which is virtually identical to \lcdm). This can also be seen from table\,\ref{tab:params3c_bao} where we compare the inferred parameters values from Planck-only and Planck+BAO data. Therefore with the inclusion of the BAO likelihood, the relative significance of the SI mode drops compared to the MI mode which can be seen from figure~\ref{fig:baopos}. The decrease in relative significance for SI mode is highest in \three{} and lowest in \one{}. This is because the increase in $ H_0 $ is the least in \one{} among all three scenarios. The $ H(z) $ and $ D_M(z) $ evolutions of the SI mode in \one{} differ the least from the corresponding MI mode evolutions and are in better agreement with the BAO data as shown in figure~\ref{fig:BAO}. Thus, upon inclusion of the BAO data, the relative significance of the SI peaks is affected the least.

%
	
	\begin{table}[t]
		\centering
		\caption{Parameter values and 68\% confidence limits in \three. }\label{tab:params3c_bao}
		\begin{tabular}{c|M{2.68cm}M{2.68cm}|M{2.68cm}M{2.68cm}}
			\hline\hline
			Parameters & \multicolumn{2}{c}{TTTEEE+lowE+lens}\vline & \multicolumn{2}{c}{TTTEEE+lowE+lens+BAO} \\
			\hline
			& SI & MI & SI & MI\\\cline{2-5}
			$\omb$ & $0.022\pm 0.00016$ & $0.022\pm 0.00014$ & $0.022\pm 0.00015$ & $0.022\pm 0.00013$\\[1ex]
			$\omc$ & $0.1204\pm 0.0013$ & $0.1199\pm 0.001$ & $0.1213\pm 0.0011$ & $0.1198\pm 0.0009$\\[1ex]
			$100\theta_s$ & $1.0463\pm 0.00094$ & $1.0419\pm 0.0003$ & $1.0461\pm 0.0011$ & $1.042\pm 0.0003$\\[1ex]
			$\ln(10^{10}A_s)$ & $2.98\pm 0.0162$ & $3.041\pm 0.0145$ & $2.981\pm 0.0163$ & $3.041\pm 0.014$\\[1ex]
			$n_s$ & $0.9388\pm 0.0043$ & $0.9629\pm 0.0048$ & $0.9362\pm 0.0038$ & $0.9633\pm 0.0044$\\[1ex]
			$\tau_\mathrm{reio}$ & $0.054\pm 0.007$ & $0.0536\pm 0.007$ & $0.051\pm 0.0069$ & $0.0535\pm 0.007$\\[1ex]
			$\lgeff$ & $-1.95\pm 0.2$ & $-4.35\pm 0.4$ & $-1.97\pm 0.24$ & $-4.36\pm 0.42$\\[1ex]
			\hline
			$H_0(\hu)$ & $69.47\pm 0.59$ & $67.9\pm 0.55$ & $68.97\pm 0.47$ & $67.96\pm 0.42$\\[1ex]
			$r_s^* (\mathrm{Mpc})$ & $144.58\pm 0.32$ & $144.88\pm 0.3$ & $144.37\pm 0.3$ & $144.91\pm 0.24$\\[1ex]
			$\sigma_8$ & $0.834\pm 0.007$ & $0.823\pm 0.006$ & $0.834\pm 0.0064$ & $0.823\pm 0.0059$\\[1ex]
			\hline
			$\chi^2 - \chi^2_{\Lambda\text{CDM}}$ & 5.45 & 0.29 & 7.45 & 0.17\\[1ex]
			\hline
		\end{tabular}
	\end{table}
    
    \subsection{Mode comparison}
    In this subsection, we analyze the relative significance of the two modes in the \sinu{} posterior. 
    We compute the Bayesian evidence $\mathcal{Z}$ for the two modes as follows
    \begin{equation}
    \mathcal{Z} \equiv \mathrm{Pr}(\mathrm{d}|M) = \int \mathrm{Pr}(\mathrm{d}|\theta,M) \mathrm{Pr}(\theta|M) \mathrm{d}\theta\,.
    \end{equation}
    Here  $M$ is the cosmological model. The Bayesian evidence essentially is the probability of the data $\mathrm{d}$ given a model $M$.
    We use the Bayesian evidence provided by \texttt{Multinest} for the two modes. Then we compute the ratio of the probabilities Pr($M|$d) of model $M$ for given data d, which is the Bayes factor, to find the relative significance of each mode assuming equal prior. This gives,
    \begin{equation}
    \mathcal{B}_\mathrm{SI} \equiv \frac{\mathrm{Pr}(M_\mathrm{SI}|\mathrm{d})}{\mathrm{Pr}(M_\mathrm{MI}|\mathrm{d})} = \frac{\mathcal{Z}_\mathrm{SI}}{\mathcal{Z}_\mathrm{MI}} \frac{\mathrm{Pr}(M_\mathrm{SI})}{\mathrm{Pr}(M_\mathrm{MI})} = \frac{\mathcal{Z}_\mathrm{SI}}{\mathcal{Z}_\mathrm{MI}}\,.
    \end{equation}
    A Bayes factor $\mathcal{B}_\mathrm{SI} > 1$, therefore, means that the SI mode is more significant than the MI mode.
    In Table \ref{tab:bayes_factor}, we show the Bayes factors for \three{}, \two{}, and \one{} for all datasets.

	\section{Discussion \& Outlook}\label{sec:discussion}
	        	\begin{table}
		\centering
		\caption{Bayes factor $\mathcal{B}_{\rm SI}$.}\label{tab:bayes_factor}
		\vspace{0.2cm}
		\begin{tabular}{lM{2cm}M{2cm}M{2cm}M{2cm}}
			\hline\hline
			Dataset & \three & \two & \one\\[1ex]\hline
			TT+lowE & 0.109 & 0.477 & 0.869\\[1ex]
			TTTEEE+lowE & 0.029 & 0.17 & 0.629\\[1ex]
			TTTEEE+lowE+lens & 0.026 & 0.164 & 0.534\\[1ex]
			TTTEEE+lowE+lens+BAO & 0.008 & 0.124 & 0.555 \\[1ex]
			TTTEEE+lowE+lens+BAO+$H_0$ & 0.137 & 0.597 & 1.147\\[1ex]
			\hline
		\end{tabular}
	\end{table}
	Secret self-interactions among the active neutrinos could leave an observable imprint in the CMB anisotropy power spectra. Such interaction stops the neutrinos from free-streaming much later than weak decoupling . As a result, the CMB power spectra experience a \emph{phase shift} and an \emph{enhancement} for the modes which entered the Hubble horizon before the decoupling of the neutrinos. We studied these effects in three cases- \three{}, \two{}, and \one{} with three, two, and one self-interacting massless neutrino species, respectively, and their cosmological implications using the latest CMB data from Planck 2018, BAO data, and local Hubble measurement  data from SH0ES. This study of the flavor-specific scenario is inspired by the recent strong constraints from several laboratory experiments on the flavor universal scenario. We make the following key observations in this work.
	\begin{itemize}
		\item In all three cases and all datasets, the posterior of $\lgeff$ has two distinct modes, namely, the SI mode characterized by a larger $\lgeff$, and the MI mode with a smaller value. The origin of the SI mode is explained as a result of degeneracies between $\geff$ and other \lcdm{} parameters. 
		The absence of free streaming neutrinos in \sinu{} leads to a phase shift and enhancement of the CMB angular power spectra. These changes are compensated by the other \lcdm{} parameters. Specifically,
		 $H_0$ compensates for the phase shift, $A_s$ and $\tau_\mathrm{reio}$ correct the overall amplitude, and $n_s$ tilts the whole spectrum to achieve a good fit to the data. We showed that this compensation mechanism works only for those values of $\geff$ for which neutrino decoupling happens close to matter radiation equality. 
		 
		 \item In intermediate values in the `valley' between the SI and MI modes are not favored by data. The large $\geff$ corresponding to the SI mode globally affects the whole CMB spectrum observed by Planck, which can be undone by changing other parameters. In contrast, the MI mode spectrum is virtually indistinguishable from the \lcdm{} as it only affects very high-$\ell$ modes, due to small $\geff$, which are not observed by Planck. The valley in between corresponds to intermediate $\geff$ values which modifies only the high-$\ell$ portion of the spectrum. These partial changes are difficult to compensate using other parameters, resulting to a worse fir to the data.
		
		\item In the \three{} scenario, the significance of the SI mode is greatly diminished if CMB polarization data is included. This is a consequence of the relatively poorer fit of the model to the low-$l$ polarization data. Even though the phase shift from the free-streaming neutrinos is the same for both temperature and polarization spectra, the peaks in the polarization spectrum are relatively sharper. As a result, the EE data is more sensitive to any change in the number of free-streaming neutrinos than the TT data\,\cite{Baumann:2015rya}. The inclusion of the polarization data also shifts the whole posteriors towards smaller $\geff$.
		
		\item However, the SI mode is rejuvenated in the flavor-specific \two{} and \one{} scenarios which are in lesser conflict with laboratory experiments. We showed that even with the polarization data, the SI mode significance is comparable or sometimes even greater than that of the MI mode. In these cases, the number of self-interacting neutrinos are less than \three{}, and as a result, the changes in the CMB spectra are also relatively moderate. This allows for more freedom to use other degenerate parameters to achieve a significantly better fit. 
		\item 
		Due to the phase shift in CMB spectrum, \sinu{} scenario favors a larger $H_0$ in the SI mode. For example, in the \three{} scenario, we find  $H_0 = 69.52 \pm 0.41 \hu$ for Planck temperature, polarization, and lensing data. Although in flavor-specific scenario, the significance of the SI mode increases, the value of $H_0$ slightly decreases due to lesser phase-shift. We do not find any strong correlation between $\sigma_8$ and $\geff$.
	\end{itemize}

	In this work, we have confined our discussion within the standard framework of only three active neutrino states. However, additional sterile neutrino states are interesting to consider for various different reasons. The presence of such an additional neutrino state(s) might change our results significantly. To begin with, the relativistic degrees of freedom $\neff$ is expected to increase, which would in turn increase the value of $H_0$~\cite{Kreisch:2019yzn}. As hinted by the short-baseline neutrino oscillation experiments, the mass of the additional sterile state is predicted to be around $\sim 1\,\mathrm{eV}$. 
	In appendix\,\ref{sec:app3}, we briefly discuss the results when extra radiation ($\dNeff$) is added. This is motivated by the sterile neutrino models. In this case, it is possible that only the sterile state can have large self-interaction.
	This scenario will evade all laboratory constraints on the active sector. However, as can be seen from table\,\ref{tab:1c2fNeff}, we do not find a large enough value of $H_0$ that can reconcile its value with the low redshift measurement data.
	Also, in order to a more careful CMB analysis with sterile neutrinos, the mass of the sterile state needs to be taken into account which could be of the order of the recombination temperature for $\sim 1\ev$ sterile neutrino.
	 The finite neutrino mass will have further implications for our analysis. In presence of massive neutrinos, the flavor-structure of the interaction matrix would certainly be non-diagonal which can have interesting signatures. Finite neutrino mass will also affect the matter power spectrum at late times.
 The strong neutrino self interaction will also enhance the primordial CMB B-mode at small scale which can have interesting implications for future B-mode experiments~\cite{Ghosh:2017jdy,Ghosh:2019tab}.

   In this work, we have kept the helium fraction $Y_p$ fixed to its BBN value. However, there is a well-known degeneracy between $Y_p$ and the number of free-streaming neutrinos. We plan to investigate it in a future work. 
   Note that, the present and all previous analyses on this topic have assumed a diagonal coupling matrix between different neutrino flavors.
    However, this is far from a realistic scenario as off-diagonal couplings will be present even in the most simple model of neutrino self-interaction. The evolution of the cosmological perturbations become much more complicated when off-diagonal couplings are present, and a more general set of evolution equations need to be used\,\cite{Oldengott:2017fhy}.
   
   Future CMB observations like, CMB-S4, will make measurements at smaller scales and will probe even higher $\ell$ values than Planck. This would probe the neutrino interaction at even earlier times, and perhaps could shed light on the nature of the mediator particle $\phi$. This work is the first step towards studying a more general and realistic flavor profile of neutrino self interaction. A more rigorous analysis including CMB, BBN, and laboratory experiment data would be worthwhile.
	
	
	\section*{Acknowledgment}
	The authors thank Rishi Khatri and Benjamin Wallisch for useful discussions. AD was supported by the U.S. Department of Energy under contract number DE-AC02-76SF00515. SG was supported in part by the National Science Foundation under Grant Number PHY-2014165 and PHY-1820860. SG also acknowledges the support from Department of Atomic Energy, Government of India during the initial stage of this project. This work used the computational facility of Department
   of Theoretical Physics, Tata Institute of Fundamental Research. 
	\appendix
	
	\section{Parameter values for other datasets}
	Here we show the parameter values and their 68\% confidence limits for the TTTEEE+lowE+lensing and TTTEEE+lowE+lensing+BAO+$H_0$ datasets. 
		\begin{table}[b]
			\centering
			\caption{Parameter values and 68\% confidence limits in \three.}\label{tab:params3c_app}
			\begin{tabular}{c|M{2.68cm}M{2.68cm}}
				\hline\hline
				Parameters & \multicolumn{2}{c}{TTTEEE+lowE+lens+BAO+$H_0$} \\
				\hline
				& SI & MI\\\cline{2-3}
				$\omb$ & $0.023\pm 0.00014$ & $0.022\pm 0.00013$\\[1ex]
				$\omc$ & $0.1206\pm 0.001$ & $0.1188\pm 0.0009$\\[1ex]
				$100\theta_s$ & $1.0465\pm 0.00079$ & $1.042\pm 0.00029$\\[1ex]
				$\ln(10^{10}A_s)$ & $2.98\pm 0.0153$ & $3.044\pm 0.0144$\\[1ex]
				$n_s$ & $0.9383\pm 0.004$ & $0.966\pm 0.0045$\\[1ex]
				$\tau_\mathrm{reio}$ & $0.0532\pm 0.007$ & $0.0563\pm 0.0071$\\[1ex]
				$\lgeff$ & $-1.91\pm 0.16$ & $-4.34\pm 0.43$\\[1ex]
				\hline
				$H_0(\hu)$ & $69.45\pm 0.42$ & $68.46\pm 0.41$\\[1ex]
				$r_s^* (\mathrm{Mpc})$ & $144.5\pm 0.26$ & $145.12\pm 0.24$\\[1ex]
				$\sigma_8$ & $0.833\pm 0.0065$ & $0.821\pm 0.006$\\[1ex]
				\hline
				$\chi^2 - \chi^2_{\Lambda\text{CDM}}$ & 1.99 & 0.17\\[1ex]
				\hline
			\end{tabular}
		\end{table}

		\begin{table}[t]
			\centering
			\caption{Parameter values and 68\% confidence limits in \two.}\label{tab:params2c_app}
			\begin{tabular}{c|M{2.68cm}M{2.68cm}}
				\hline\hline
				Parameters & \multicolumn{2}{c}{TTTEEE+lowE+lens+BAO+$H_0$} \\
				\hline
				& SI & MI\\\cline{2-3}
				$\omb$ & $0.022\pm 0.0001$ & $0.022\pm 0.00013$\\[1ex]
				$\omc$ & $0.12\pm 0.001$ & $0.1188\pm 0.0009$\\[1ex]
				$100\theta_s$ & $1.045\pm 0.00068$ & $1.042\pm 0.00029$\\[1ex]
				$\ln(10^{10}A_s)$ & $3.0\pm 0.0151$ & $3.044\pm 0.0145$\\[1ex]
				$n_s$ & $0.9483\pm 0.004$ & $0.966\pm 0.0046$\\[1ex]
				$\tau_\mathrm{reio}$ & $0.0544\pm 0.007$ & $0.0565\pm 0.0071$\\[1ex]
				$\lgeff$ & $-1.91\pm 0.22$ & $-4.22\pm 0.52$\\[1ex]
				\hline
				$H_0(\hu)$ & $69.08\pm 0.42$ & $68.47\pm 0.4$\\[1ex]
				$r_s^* (\mathrm{Mpc})$ & $144.73\pm 0.26$ & $145.12\pm 0.23$\\[1ex]
				$\sigma_8$ & $0.827\pm 0.0065$ & $0.821\pm 0.0059$\\[1ex]
				\hline
				$\chi^2 - \chi^2_{\Lambda\text{CDM}}$ & -1.35 & 0.25\\[1ex]
				\hline
			\end{tabular}
		\end{table}
	
	   \begin{table}[t]
	   	\centering
	   	\caption{Parameter values and 68\% confidence limits in \one.}\label{tab:params1c_app}
	   	\begin{tabular}{c|M{2.68cm}M{2.68cm}}
	   		\hline\hline
	   		Parameters & \multicolumn{2}{c}{TTTEEE+lowE+lens+BAO+$H_0$} \\
	   		\hline
	   		& SI & MI\\\cline{2-3}
	   		$\omb$ & $0.022\pm 0.0001$ & $0.022\pm 0.00013$\\[1ex]
	   		$\omc$ & $0.12\pm 0.0009$ & $0.1188\pm 0.0009$\\[1ex]
	   		$100\theta_s$ & $1.043\pm 0.00056$ & $1.042\pm 0.00029$\\[1ex]
	   		$\ln(10^{10}A_s)$ & $3.0\pm 0.0151$ & $3.045\pm 0.0142$\\[1ex]
	   		$n_s$ & $0.9572\pm 0.004$ & $0.966\pm 0.0042$\\[1ex]
	   		$\tau_\mathrm{reio}$ & $0.0554\pm 0.007$ & $0.0566\pm 0.0071$\\[1ex]
	   		$\lgeff$ & $-1.86\pm 0.36$ & $-4.03\pm 0.61$\\[1ex]
	   		\hline
	   		$H_0(\hu)$ & $68.75\pm 0.41$ & $68.48\pm 0.41$\\[1ex]
	   		$r_s^* (\mathrm{Mpc})$ & $144.93\pm 0.24$& $145.12\pm 0.24$\\[1ex]
	   		$\sigma_8$ & $0.822\pm 0.0071$ & $0.821\pm 0.006$\\[1ex]
	   		\hline
	   		$\chi^2 - \chi^2_{\Lambda\text{CDM}}$ & -1.67 & 0.33\\[1ex]
	   		\hline
	   	\end{tabular}
	   \end{table}

	\section{Posterior distributions of all parameters}
	\begin{figure}
		\centering
		\includegraphics[width=\linewidth]{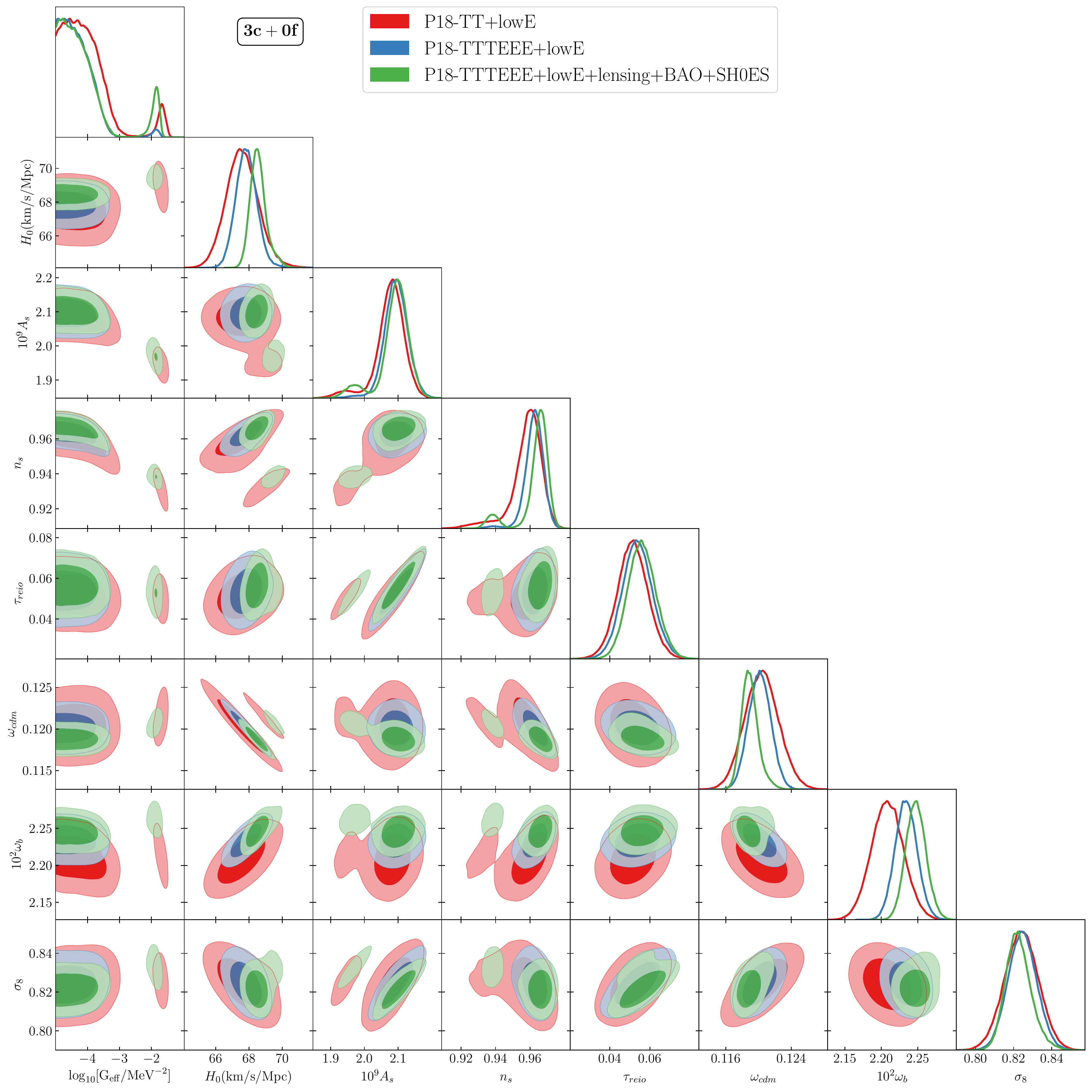}
		\caption{The 68\% and 95\% confidence limits of all parameters in \three{} for the Planck 2018 TT+lowE, TTTEEE+lowE, and TTTEEE+lowE+lensing+BAO+$H_0$ datasets.}
		\label{fig:3c-tri-pos}
	\end{figure}
	\begin{figure}
		\centering
		\includegraphics[width=\linewidth]{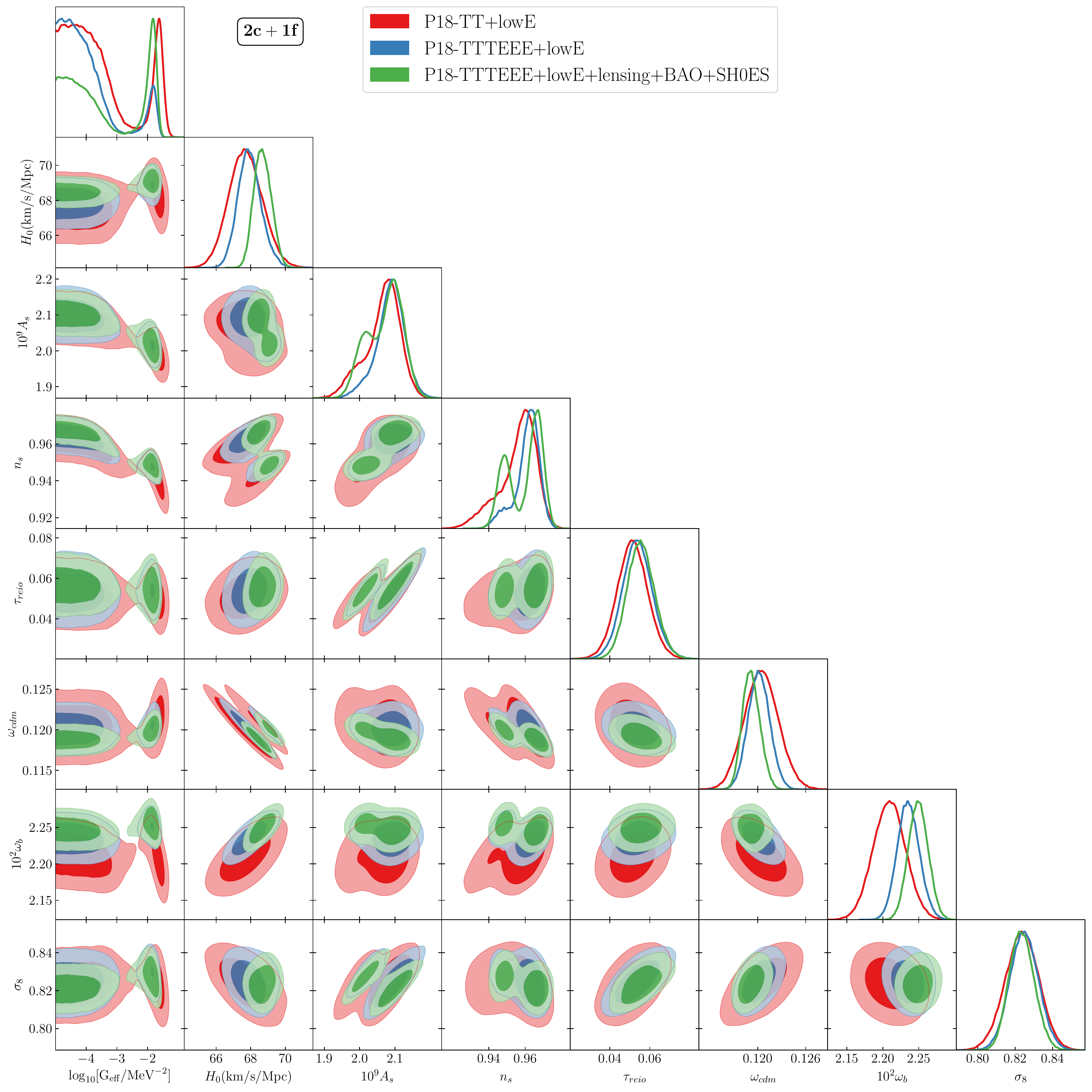}
		\caption{The 68\% and 95\% confidence limits of all parameters in \two{} for the Planck 2018 TT+lowE, TTTEEE+lowE, and TTTEEE+lowE+lensing+BAO+$H_0$ datasets.}
		\label{fig:2c-tri-pos}
	\end{figure}
	\begin{figure}
		\centering
		\includegraphics[width=\linewidth]{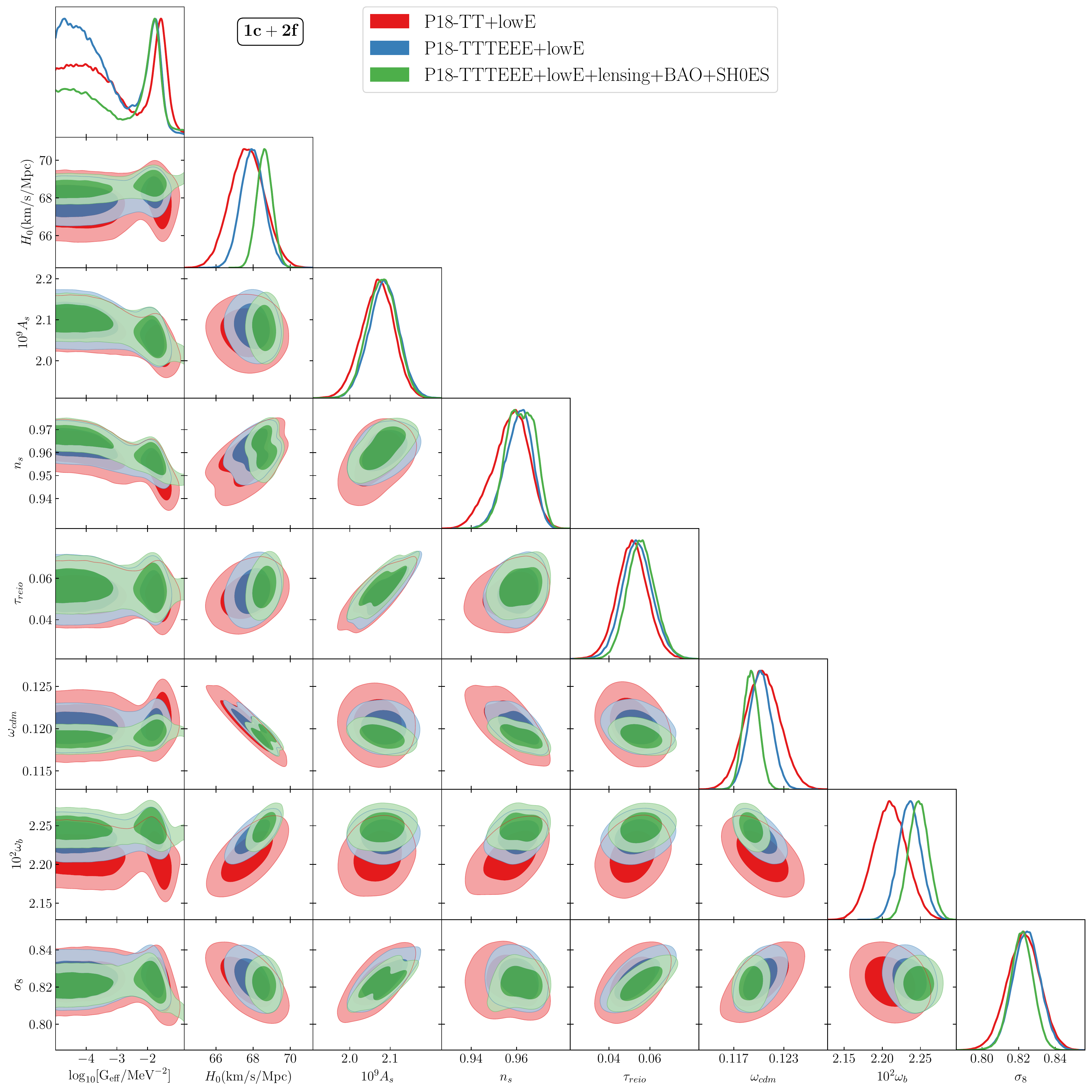}
		\caption{The 68\% and 95\% confidence limits of all parameters in \one{} for the Planck 2018 TT+lowE, TTTEEE+lowE, and TTTEEE+lowE+lensing+BAO+$H_0$ datasets.}
		\label{fig:1c-tri-pos}
	\end{figure}

\section{Flavor dependent SINU with additional radiation species}\label{sec:app3}
\begin{figure}
	\centering
	\includegraphics[width=\linewidth]{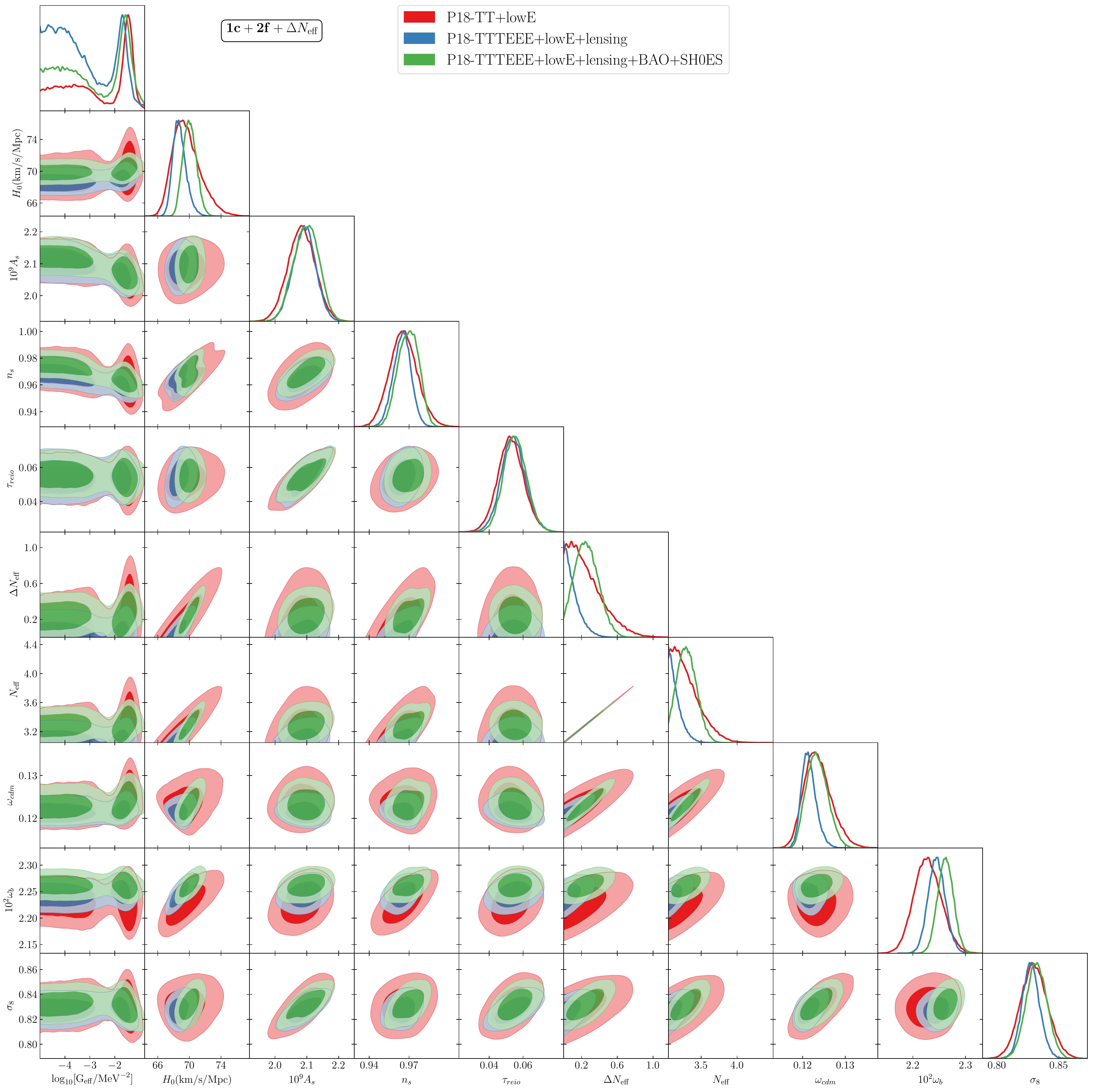}
	\caption{The 68\% and 95\% confidence limits of all parameters in \one{} + $ \Delta N_{\rm eff} $ for the Planck 2018 TT+lowE, TTTEEE+lowE, and TTTEEE+lowE+lensing+BAO+$H_0$ datasets.}
	\label{fig:1c-Neff-tri-pos}
\end{figure}

In this section, we briefly discuss the effects of SINU cosmology with varying $ N_{\rm eff} $. Flavor universal SINU with additional interacting radiation species are interesting as they can accommodate even higher value of the Hubble constant. For this analysis, we chose the scenario \one{} +  $ \Delta N_{\rm eff} $ which contains additional free streaming radiation with effective neutrino number $ \Delta N_{\rm eff} $. One possible model interpretation of this setup is the existence of sterile neutrino state(s) among which one is self-interacting, while all other neutrinos are free-streaming.
 Figure \ref{fig:1c-Neff-tri-pos} shows the $ 1$-D and $ 2 $-D posteriors of all the parameters. The value of $ H_0 $ increase slightly compared to the \one{} scenario due to presence of the additional radiation relics which can be seen from Table~\ref{tab:1c2fNeff}.

\renewcommand{\arraystretch}{1.4}
\begin{table}\label{tab:app3}
	\centering
	\caption{values of $ H_0 $ ($ 1\sigma $ errorbar) and upper limit of $ N_{\rm eff} $ ($ 95\% $ C.L) for \one{} +  $ \Delta N_{\rm eff} $ for all dataset.}
	\begin{tabular}{l|c|c|c}
		\hline
		\hline
		Parameters & TT+lowE & TTTEEE+lowE+lensing & TTTEEE+lowE+lensing+BAO+H0\\
		\hline
		$ H_0 $ & $ 69.7^{+1.3}_{-2.1} $ & $ 68.77^{+0.66}_{-0.95} $ & $ 70.04^{+0.84}_{-0.84} $\\
		\hline
		$ N_{\rm eff} $ & $ <3.76 $ & $ <3.38 $ & $ <3.58 $\\
		\hline
	\end{tabular}
	\label{tab:1c2fNeff}
\end{table}

%
%

	\bibliographystyle{jhep}
	\bibliography{sinu}
\end{document}